\documentclass[twocolumn,aps,prb,reprint,showpacs]{revtex4-1}

\pdfoutput=1

\usepackage{units}
\usepackage{bm}
\usepackage{amsmath}
\usepackage{amssymb}
\usepackage{graphicx}
\usepackage{natbib}
\usepackage{hyperref}
\makeatletter

\@ifundefined{textcolor}{}
{%
\definecolor{BLACK}{gray}{0}
\definecolor{WHITE}{gray}{1}
\definecolor{RED}{rgb}{1,0,0}
\definecolor{GREEN}{rgb}{0,1,0}
\definecolor{BLUE}{rgb}{0,0,1}
\definecolor{CYAN}{cmyk}{1,0,0,0}
\definecolor{MAGENTA}{cmyk}{0,1,0,0}
\definecolor{YELLOW}{cmyk}{0,0,1,0}
}

\@ifundefined{textcolor}{}{
\definecolor{BLACK}{gray}{0}
\definecolor{WHITE}{gray}{1}
\definecolor{RED}{rgb}{1,0,0}
\definecolor{GREEN}{rgb}{0,1,0}
\definecolor{BLUE}{rgb}{0,0,1}
\definecolor{CYAN}{cmyk}{1,0,0,0}
\definecolor{MAGENTA}{cmyk}{0,1,0,0}
\definecolor{YELLOW}{cmyk}{0,0,1,0}
}

\usepackage{xcolor}

\newcommand{\eq}[1]{Eq.~(\ref{#1})}

\newcommand{\beq}{\begin{equation}}
\newcommand{\eeq}{\end{equation}}

\makeatother

\begin{document}

\title{Substrate-induced Majorana renormalization in topological nanowires}

\author{S. Das Sarma}

\affiliation{Department of Physics, Condensed Matter Theory Center and Joint Quantum
Institute, University of Maryland, College Park, Maryland 20742-4111,
USA}

\author{Hoi-Yin Hui}

\affiliation{Department of Physics, Condensed Matter Theory Center and Joint Quantum
Institute, University of Maryland, College Park, Maryland 20742-4111,
USA}

\author{P. M. R. Brydon}

\affiliation{Department of Physics, Condensed Matter Theory Center and Joint Quantum
Institute, University of Maryland, College Park, Maryland 20742-4111,
USA}

\author{Jay D. Sau}

\affiliation{Department of Physics, Condensed Matter Theory Center and Joint Quantum
Institute, University of Maryland, College Park, Maryland 20742-4111,
USA}

\date{\today}
\begin{abstract}
We theoretically consider the substrate-induced Majorana localization
length renormalization in nanowires in contact with a bulk superconductor
in the strong tunnel-coupled regime, showing explicitly that this
renormalization depends strongly on the transverse size of the one-dimensional
nanowires. For metallic (e.g. Fe on Pb) or semiconducting (e.g. InSb
on Nb) nanowires, the renormalization effect is found to be very strong
and weak respectively because the transverse confinement size in the
two situations happens to be 0.5~nm (metallic nanowire) and 20~nm
(semiconducting nanowire). Thus, the Majorana localization length
could be very short (long) for metallic (semiconducting) nanowires
even for the same values of all other parameters (except for the transverse
wire size). We also show that any tunneling conductance measurements
in such nanowires, carried out at temperatures and/or energy resolutions
comparable to the induced superconducting energy gap, cannot distinguish
between the existence of the Majorana modes or ordinary subgap fermionic
states since both produce very similar broad and weak peaks in the
subgap tunneling conductance independent of the localization length
involved. Only low temperature (and high resolution) tunneling measurements
manifesting sharp zero bias peaks can be considered to be signatures
of Majorana modes in topological nanowires.
\end{abstract}
\maketitle

\section{Introduction}

Majorana fermions (MF), which were proposed theoretically 80 years
ago as real solutions of the Dirac equation in the context of understanding
neutrinos \cite{Majorana1937}, have recently found their incarnations
in solid state systems \cite{Read2000Paired,Kitaev2001Unpaired,Fu2008Superconducting,Sau2010Generic}
as zero-energy localized excitations in topological superconductors
(TS). In addition to the defining property of being their own anti-particles,
MFs in solid state systems are known to have exotic non-Abelian braiding
statistics \cite{Read2000Paired}, making them of high theoretical
interest with a potential application in topological quantum computation
\cite{Nayak2008Non-Abelian,DasSarma2015Majorana}. While the initial
proposals involving exotic materials (e.g. $p$-wave superconductors
\cite{Read2000Paired,Kitaev2001Unpaired}, $\nicefrac{5}{2}$-fractional
quantum Hall states \cite{Read2000Paired}, and topological insulators
\cite{Fu2008Superconducting}) have so far escaped experimental realization,
more recent proposals utilizing spin-orbit-coupled semiconductors
\cite{Sau2010Generic,Sau2010Non-Abelian,Lutchyn2010Majorana,Oreg2010Helical}
or magnetic adatoms \cite{Chung2011Topological,Duckheim2011Andreev,Choy2011Majorana,Nadj-Perge2013Proposal}
proximity-coupled to $s$-wave superconductors have been the major
focus of experimental efforts, with several published works claiming
the observation of signatures of MFs in such systems \cite{Mourik2012Signatures,Deng2012Anomalous,Rokhinson2012fractional,Das2012Zero-bias,Churchill2013Superconductor-nanowire,Finck2013Anomalous,Nadj-Perge2014Observation}.
The initial claim of the observation of {}``Signatures of Majorana
Fermions in Hybrid Superconductor-Semiconductor Nanowire Devices''
by Mourik \textit{et~al.} \cite{Mourik2012Signatures} in the InSb/Nb
system, following precise theoretical predictions \cite{Sau2010Non-Abelian,Lutchyn2010Majorana,Oreg2010Helical},
was later experimentally replicated by several groups in both InSb/Nb
\cite{Mourik2012Signatures,Deng2012Anomalous,Rokhinson2012fractional,Churchill2013Superconductor-nanowire}
and InAs/Al \cite{Das2012Zero-bias,Finck2013Anomalous} hybrid structures,
giving considerable confidence in the universal nature of the underlying
physical phenomena. Despite this apparent experimental success,
the identification of the zero-bias peak as a MF is still
debated~\cite{BagretsPRL2012,*PikulinNJP2012}; more definitive proof, such as signatures of the  predicted
non-Abelian braiding properties, is yet to be detected as unambiguous
evidence for the existence of the Majorana mode~\cite{Nayak2008Non-Abelian,DasSarma2015Majorana}.

The motivation underlying the semiconductor-superconductor heterostructure
Majorana platform \cite{Sau2010Non-Abelian} is the artificial creation
of a spinless low-dimensional (either 2D \cite{Sau2010Generic,Sau2010Non-Abelian}
or 1D \cite{Sau2010Non-Abelian,Lutchyn2010Majorana,Oreg2010Helical})
$p$-wave superconductor supporting MFs \cite{Kitaev2001Unpaired}.
The effectively spinless $p$-wave superconductivity residing in the
semiconductor serves as the TS here arising from a combination of
spin-orbit coupling, spin-splitting, and ordinary $s$-wave superconductivity.
The combination of spin-splitting and spin-orbit coupling in the semiconductor
allows, under appropriate conditions (of large enough spin-splitting
and spin-orbit coupling), ordinary $s$-wave singlet Cooper pairs
to tunnel from the superconductor to the semiconductor enabling topological
$p$-wave superconductivity with triplet superconducting correlations
\cite{Liu2015Universal} to develop in the semiconductor through proximity
coupling. The experimentally relevant topological system has been
an InSb \cite{Mourik2012Signatures,Deng2012Anomalous,Rokhinson2012fractional,Churchill2013Superconductor-nanowire}
or InAs \cite{Das2012Zero-bias,Finck2013Anomalous} nanowire on a
Nb or Al superconducting substrate with the Zeeman spin splitting
achieved through the application of an external magnetic field. For
the magnetic field larger than a critical value, which is given simply
by the proximity-induced superconducting gap in the nanowire (assuming
the chemical potential in the nanowire can be taken to be zero), the
nanowire becomes an effective TS with zero-energy MFs localized at
the wire ends \cite{Kitaev2001Unpaired,Sau2010Non-Abelian,Lutchyn2010Majorana,Oreg2010Helical}.
These zero-energy MFs should lead to zero-bias conductance peaks (ZBCP)
in the tunneling conductance measurement \cite{Sau2010Non-Abelian},
and the experimental observation \cite{Mourik2012Signatures,Churchill2013Superconductor-nanowire,Das2012Zero-bias,Finck2013Anomalous}
of such field-induced ZBCP has been taken as evidence for the existence
of MFs in these topological nanowires.

Typically, these MFs are localized near the ends of the wire with
a finite Majorana localization length $\xi$ which equals the superconducting
coherence length in the topological nanowire. When the wire length
$L\gg\xi$, the two MFs at the two ends of the wire are considered
to be in the topologically exponential protection regime with the
MF wavefunction falling off as $e^{-x/\xi}$ along the wire (modulo
some oscillations not of particular interest here \cite{Cheng2009Splitting,Cheng2010Stable,DasSarma2012Splitting}).
The MF is a well-defined zero-energy non-Abelian mode only in this
exponentially protected ($L\gg\xi$) regime. By contrast, for short
wires or long coherence length (i.e. $L<\xi$), the two end MFs hybridize,
producing split peaks shifted away from zero energy (and these split
peaks represent fermionic subgap states rather than MFs), and the
topological protection no longer applies. It is only when the Majorana
splitting is exponentially small (i.e. $L\gg\xi$), the nanowire system
can be considered to be topological \cite{DasSarma2015Majorana}.
Thus, the quantitative magnitude of $\xi$ (or more precisely the
dimensionless length ratio $L/\xi$) is a key ingredient in the physics
of MFs. We note that the localized zero-energy MF bound states are
also often called the Majorana modes, and we use the terminology Majorana
fermions and Majorana modes interchangeably in this paper to mean
the same zero-energy MF subgap localized TS excitations in a spinless
p-wave superconductor. The current experimental MF search is mostly
focused on looking for subgap zero-bias tunneling conductance peaks
associated with these localized zero-energy Majorana modes -- ideally,
the ZBCP should have the quantized value of $2e^{2}/h$, but experimentally
the ZBCPs observed so far have actual conductance values factors of
10 ($10^{4}$) lower in semiconductor nanowires \cite{Mourik2012Signatures,Das2012Zero-bias,Churchill2013Superconductor-nanowire}
(ferromagnetic nanowires \cite{Nadj-Perge2014Observation}). Finite
temperature, finite wire length, finite tunnel barrier, finite experimental
resolution, unwanted fermionic subgap states, and possible inelastic
processes conspire together to suppress the experimental ZBCP strength,
and this non-observation of perfect ZBCP quantization, which is much
more severe in the ferromagnetic nanowires than in semiconductor systems,
remains an open question in the subject.

The MF localization length (or equivalently the TS coherence length)
$\xi$ is often assumed to be given by the standard superconducting
coherence length formula, $\xi\sim v_{F}/\Delta$, where $v_{F}$
and $\Delta$ are respectively the Fermi velocity and the induced
TS gap in the nanowire. This superconducting coherence length formula
is certainly appropriate for MF localization if the nanowire can be
considered an isolated spinless $p$-wave superconductor with the
Majorana modes localized at the wire ends (which serve as the defects
localizing the MFs). But in the experimentally relevant situation
the nanowire is not isolated, it is in fact in contact with (or lying
on top of) a superconducting substrate which provides the necessary
proximity effect to produce the topological system in conjunction
with spin-orbit coupling and spin splitting. The question therefore
arises whether the MF localization length formula is modified from
the simple coherence length formula, or equivalently, whether the
Fermi velocity and/or the appropriate nanowire superconducting gap
are renormalized by the substrate. This issue was in fact discussed
by Sau \textit{et~al.} \cite{Sau2010Robustness} and Stanescu \textit{et~al.}
\cite{Stanescu2010Proximity} some years ago in the context of 2D
sandwich structures involving semiconductor/superconductor and topological-insulator/superconductor
heterostructures, and very recently by Peng \textit{et~al.} \cite{Peng2014Strong}
in the context of 1D ferromagnetic nanowire on superconductor hybrid
structures used in the recent Princeton STM experiment \cite{Nadj-Perge2014Observation}.
(The actual system theoretically considered by Peng \textit{et~al.}
\cite{Peng2014Strong} is in fact a helical magnetic adatom chain,
not a ferromagnetic chain, on a superconducting substrate.) In the
first part of the current work, we theoretically study the MF
localization question 
in depth for 1D topological nanowire hybrid systems, discussing the
substrate-induced MF renormalization for both semiconductor and ferromagnetic
nanowires on an equal footing, comparing and contrasting the two
situations. In particular, we address the important issue of how
it might be possible that Majorana localization length could be
strongly (weakly) renormalized in ferromagnetic (semiconductor)
nanowire systems studied in different laboratories.

Recently, it has been proposed that the three ingredients for the
semiconductor nanowire proposal, i.e. superconductivity, magnetization,
and spin-orbit coupling, can be realized in ferromagnetic nanowires
deposited on a spin-orbit coupled superconductor. Experimental evidence
in the form of a weak and broad zero-bias peak seems to provide some
support to this hypothesis \cite{Nadj-Perge2014Observation}. Several
theoretical calculations \cite{Nadj-Perge2013Proposal,Hui2015Majorana,Dumitrescu2014Majorana,Li2014Topological,HeimesNJP2015}
have shown that as a matter of principle Majorana modes can emerge
in ferromagnetic wires in superconductors, as had been suggested in
more mesoscopic geometries \cite{Chung2011Topological,Duckheim2011Andreev,Takei2012Microscopic,Wang2010Interplay}.
Motivated by earlier STM works \cite{Yazdani1997Probing,Ji2008High-Resolution},
some of the theoretical works \cite{Choy2011Majorana,Pientka2013Topological,Kim2014Helical,HeimesNJP2015}
have modeled the system to be a chain of magnetic atoms {[}i.e.
Yu-Shiba-Rusinov (YSR) impurities{]} on the superconductor surface,
with no direct hopping between the impurity orbitals. This class of
proposals supports MFs only in a limited parameter regime \cite{Brydon2015Topological,Peng2014Strong}.
Although the YSR limit and the ferromagnetic nanowire limit are the
two extreme crossover regimes (tuned by very weak and very strong
inter-site hopping in the nanowire, respectively) of the same underlying
Hamiltonian (i.e. there is no quantum phase transition separating
them, it is simply a hopping-induced crossover from the YSR regime
to the ferromagnetic wire regime as hopping increases), the TS properties
in the two limits are very different. In the YSR limit, considerable
fine-tuning of the chemical potential is necessary in order to achieve
TS and MF \cite{Brydon2015Topological,HeimesNJP2015}, whereas the TS
with localized 
MF arise generically without any fine-tuning in the ferromagnetic
nanowire limit of strong hopping. Thus, any possible generic existence
of MF in the magnetic adatom chain on superconducting substrates is
more natural in the ferromagnetic wire limit \cite{Hui2015Majorana,Dumitrescu2014Majorana}
rather than in the YSR limit. Therefore, considering the system as
a ferromagnetic wire in proximity to a superconductor is the natural
way to understand the zero-bias conductance. This puts the ferromagnetic
chain and the semiconductor nanowire topological systems on an equal
footing with the only difference being that in the semiconductor wire
(the ferromagnetic chain) case the spin-splitting arises from an externally
applied magnetic field (an intrinsic ferromagnetic exchange splitting).
However, in the semiconductor nanowire proposals \cite{Lutchyn2010Majorana,Oreg2010Helical}
the decay length of the Majorana is typically found to be comparable
to the bulk coherence length in the superconductor. This also seems
to be a common feature in the simulations of the ferromagnetic nanowire
systems so far since the substrate-induced renormalization of the
nanowire parameters is not included in the theory, thus considering
the nanowire to be effectively isolated~\cite{Nadj-Perge2013Proposal,Hui2015Majorana,Dumitrescu2014Majorana}.
On the other hand, it was noted that YSR bound states in STM experiments
appeared to show a much shorter decay length than the coherence length
\cite{Yazdani1997Probing,Ji2008High-Resolution}. Based on this, it
was conjectured \cite{Dumitrescu2014Majorana} that the Majorana modes
might appear to be confined to length-scales shorter than the coherence
length because of the delocalization of the wavefunction into the
bulk superconductor. Very recent work using a spin-helical adatom chain
model~\cite{Peng2014Strong} has shown how this substrate-induced 
renormalization mechanism may suppress the coherence length, possibly
leading to a short MF localization length in a magnetic
chain on a superconductor which is very strongly tunnel-coupled to
the magnetic chain, even if the topological superconducting gap is
very small. Anomalously short Majorana localization lengths were
numerically demonstrated for similar models
in~Refs.~\cite{Li2014Topological} and~\cite{HeimesNJP2015}. Whether
such a scenario applies to the actual experimental situation of
Ref.~\cite{Nadj-Perge2014Observation} is currently 
unknown.

The actual MF localization length question is of great importance
to the experimental observations in \cite{Nadj-Perge2014Observation}
since the estimated TS gap in the Fe adatom chains on Pb substrates
studied therein is very small ($\sim0.1\,{\rm meV}$) leading to a
rather long coherence length (or equivalently MF localization length)
of $>100\,{\rm nm}$ (assuming no substrate-induced renormalization)
which would be much larger than the typical length of the adatom chains
($5-50\,{\rm nm}$) used in Ref.~\cite{Nadj-Perge2014Observation}.
In such a situation, the TS system is not in the exponentially protected
regime at all, and the two end MFs should hybridize strongly leading
to ordinary uninteresting fermionic states at high energies. 
Thus, to the extent the observations in \cite{Nadj-Perge2014Observation}
manifest MF features, one must understand how the
very long TS coherence length (i.e. the MF localization length) associated
with the small induced superconducting gap can be consistent with
the existence of isolated (i.e. non-hybridized) MFs in a system where
the wire length is shorter than the localization length. (This conceptually
problematic situation does not seem to arise in the semiconductor
TS systems \cite{Mourik2012Signatures,Deng2012Anomalous,Rokhinson2012fractional,Das2012Zero-bias,Churchill2013Superconductor-nanowire,Finck2013Anomalous}
since the wire length ($>1\,{\rm \mu m}$) is typically much larger
than the MF localization length ($\sim10\,{\rm nm}$) in the semiconductor
nanowire systems--- in fact, systematic experimental efforts appeared
to have observed the predicted Majorana hybridization effect in the
semiconductor nanowires in the regime of long MF localization length
induced by a large external magnetic field \cite{Churchill2013Superconductor-nanowire,Finck2013Anomalous}.)
If the observed subgap conductance features in Ref.~\cite{Nadj-Perge2014Observation}
are indeed implying the existence of MFs in the underlying ferromagnetic
nanowire, as has been concluded \cite{Nadj-Perge2014Observation},
then it is imperative that a clear theoretical understanding is developed
for why the TS coherence length being much larger than the topological
wire length is not a problem. One possible way out of this quandary,
suggested by Peng \textit{et~al.} \cite{Peng2014Strong} using a
helical magnetic chain model, is a strong suppression of
the MF localization length by the substrate so that the $L\gg\xi$
condition is still satisfied because the TS coherence length is renormalized
by the substrate. This solution comes with the caveat that that the
strongly localized MF is accompanied by a power-law tail~\cite{Pientka2013Topological},
which extends to the coherence length of the bulk superconductor.
So while the localization length appears to be reduced on length-scales
where the splitting is measurable in tunneling, it is not clear that
this suppression significantly aids the splitting problem for quantum
information purposes because of the power-law tail. We emphasize that
it is not clear at all at this stage that the condition (in particular,
very strong tunnel coupling between the substrate and the chain) for
such strong substrate-induced renormalization of the MF localization
length is actually operational in the experiment of Ref.~\cite{Nadj-Perge2014Observation},
but the possibility of the substrate-induced suppression of the MF
localization length must be taken seriously since it provides a way
forward for future experiments to test this hypothesis as a possible
resolution of the long TS coherence 
length conundrum, e.g., by studying the
MF hybridization or splitting systematically as a function of the
dimensionless ratio $\xi/L$ by changing $L$ in a controlled manner.

The possibility that the MF localization length is strongly suppressed
by the substrate immediately brings up the question of whether such
a substrate-induced renormalization phenomenon is also operational
in the semiconductor TS systems and, if so, the possible implications
for the semiconductor MF experiments \cite{Mourik2012Signatures,Deng2012Anomalous,Rokhinson2012fractional,Das2012Zero-bias,Churchill2013Superconductor-nanowire,Finck2013Anomalous}
which have so far been simply interpreted on the basis of the standard
$\xi\sim v_{F}/\Delta$ formula with no substrate-induced coherence
length suppression. The corresponding renormalization question for
semiconductor-superconductor 2D hybrid structures was studied in depth
in Refs.~\cite{Stanescu2010Proximity,Peng2014Strong}, and here we
generalize the theory to 1D semiconductors and ferromagnetic metals
in proximity to bulk superconductors. One possible reconciliation
for why the ferromagnetic (semiconductor) nanowire MF localization
length is strongly (not) renormalized by the substrate is simply by
assuming that the ferromagnetic adatoms (semiconductor wire) are (are
not) strongly tunnel-coupled to the substrate superconductor, but
we want to avoid such \textit{ad hoc} assumptions. One question we
address in the first part of the paper is how different the substrate-induced
Majorana localization length renormalization can be in metallic and
semiconductor TS nanowires assuming essentially identical conditions
to be prevailing for the substrate properties (including equivalently
strong tunnel coupling of the wire to the substrate) in both cases.

In this paper, we discuss in general the localization length of Majorana
modes in proximity-induced superconductors. To set a context, we start
by discussing the localization length of Majorana modes in the Kitaev
chain \cite{Kitaev2001Unpaired} in various parameter regimes finding
that depending on parameter values the localization length can vary
from being of the order of a lattice spacing to more than many hundreds
of lattice spacings as is typical for the coherence length in ordinary
superconductors. Realizations of such Kitaev chains where the Majorana
localization length is of the order of several lattice sites have
been proposed in quantum dot arrays \cite{Sau2012Realizing}, and
therefore, in principle, the variation in the MF localization properties
can be tested in the linear quantum dot arrays by suitably tuning
the dot parameters. Following this, we explicitly consider the proximity
effect of the bulk superconductor with a goal to understanding the
length-scale problem for ferromagnetic nanowires in proximity to spin-orbit
coupled superconductors as used in the experiment of Ref.~\cite{Nadj-Perge2014Observation}.
While the superconducting proximity effect is simply modeled by a
pairing potential in most papers on the subject, a more microscopic
consideration \cite{Sau2010Robustness} suggests that the proximity
effect should be represented by a non-local frequency-dependent self-energy
\begin{equation}
\Sigma\left(\omega;\bm{r},\bm{r}'\right)=T\left(\bm{r}\right)G_{SC}\left(\omega;\bm{r},\bm{r}'\right)T\left(\bm{r}'\right)^{\dagger},\label{eq:tGt}
\end{equation}
 where the matrix $T$ represents the hopping between the nanowire
and the superconductor, and $G_{SC}(\omega;\bm{r},\bm{r}')$ is the
Green function of the superconductor. By approximating the Green function
by that of a bulk $s$-wave superconductor, we show below that the
local part of the self-energy has the form of 
\begin{equation}
\Sigma\left(\omega;\bm{r},\bm{r}\right)\sim-\frac{\Gamma\omega}{\Delta_{SC}}\hat{\tau}_{0}+\Gamma\hat{\tau}_{x},\label{eq:Sigma_Intro}
\end{equation}
where $\Gamma$ is the parameter determining the strength of the
superconducting proximity coupling in the nanowire. As we argue in
Sec.~\ref{sub:Dependence-of-Gamma}, based on a microscopic derivation,
the proximity parameter $\Gamma\sim(R/a)^{-3}E_{F}$, where
  $a\sim0.5\,{\rm nm}$ and $E_{F}\sim1\,{\rm eV}$ are respectively a
  length scale on the order of the lattice constant
and the Fermi energy in the superconductor, and $R$ is the radius
of the nanowire. In mesoscopic semiconductor nanowire geometries $R\sim20{\rm nm}$
leading to $\Gamma\sim0.01{\rm meV}\ll\Delta_{SC}$, and this fits
into the simple picture for the proximity effect where retardation
effects associated with the frequency dependence in Eq.~(\ref{eq:Sigma_Intro})
may be ignored. Atomistic ferromagnetic wires \cite{Nadj-Perge2014Observation}
are qualitatively different since $R\sim0.5{\rm nm}$ for these wires
and the estimated $\Gamma\sim1{\rm eV}$. This clearly puts the analysis
in the regime $\Gamma\gg\Delta_{SC}$, which is the strongly renormalized
limit \cite{Sau2010Robustness}. Establishing this key difference
between the MFs in semiconductor and metallic nanowires (i.e. the
MF localization length is strongly renormalized in one, but not in
the other, due to substrate renormalization arising from retardation
effects in the proximity self-energy function) is a main goal of this
paper. While it might appear that a proximity effect of $\Gamma\gg\Delta_{SC}$
in the ferromagnetic nanowire system would produce an $s$-wave pairing
in the nanowire that is much greater than $\Delta_{SC}$ in contradiction
with experiment, the frequency dependence of the self-energy, where
$\Gamma$ enters as a parameter, ensures that the $s$-wave pairing
in this case is $\Delta_{0}\sim\frac{\Gamma\Delta_{SC}}{\Delta_{SC}+\Gamma}\sim\Delta_{SC}$,
which is much less than $\Gamma$. Therefore, it is clear that the
full frequency-dependent self-energy is critical to get the physics
correct \cite{Sau2010Robustness}.

We present calculations of the local density of states for the ferromagnetic
nanowire in proximity to a spin-orbit coupled superconductor. We find
that the frequency dependence introduces renormalization of all microscopic
parameters in a way which drastically reduces the coherence length
in the ferromagnetic nanowire, in contrast to the semiconductor nanowire
MFs where the standard definition for the coherence length applies
with little renormalization by the substrate. After establishing that
the MF localization in the ferromagnetic nanowire system could indeed
be very short in spite of the induced topological superconducting
gap being very small, we consider the actual experimental situation
\cite{Nadj-Perge2014Observation} where the MF observation in a Fe
chain on a superconducting Pb substrate has been claimed in an STM
study. At first, the very small induced gap ($\sim100\,{\rm \mu eV}$)
in the experiment seems to indicate a very long MF localization length
much larger than the length of the Fe chains, casting serious doubt
on the experimental interpretation of lattice-scale MF localization
in the system. Peng \textit{et~al.} \cite{Peng2014Strong}, however,
provided a way out of this puzzle by showing that, assuming the Pb-Fe
tunnel coupling is strong, it is possible for the MF localization
length to be short in spite of the induced gap being small within
their helical chain model. In the current work, we go beyond the
analysis in Ref.~\cite{Peng2014Strong} by showing that this
renormalization is a generic feature of chain TS
platforms. Moreover, we directly address the question of why such
effects are weak in the semiconductor nanowire, revealing the
key role of mesoscopic vs microscopic geometry in determining the substrate
coupling strength. This provides a clearer understanding of the 
relationship between these the superficially different chain and
nanowire TS platforms.

In the second part of the paper, we address the issue of energy 
localization of the Majorana zero mode in the recent ferromagnetic
wire STM experiment~\cite{Nadj-Perge2014Observation}. Apart from
spatial localization, another peculiar feature of the Majorana 
modes in the ferromagnetic nanowire experiment relates to the broadening
of the spectrum in the energy domain. This feature of the experiment
\cite{Nadj-Perge2014Observation} results from the high temperatures
($T=1.4\,{\rm K}$) at which the experiment is carried out comparable
to the induced gap ($\sim0.1\,{\rm meV}$) itself. Such high temperatures
limit the energy resolution of the STM, thus compromising
the claim of the MF observation in Ref.~\cite{Nadj-Perge2014Observation}.
The Majorana mode must be localized both spatially and spectrally,
i.e., the MF must be a well-defined zero energy excitation (in addition
to being spatially localized at the wire ends) which becomes problematic
at high temperatures where the zero mode hybridizes with the above-gap
fermionic particle-hole excitations, thus completely losing its non-Abelian
Majorana character. The energy localization property is independent
of the spatial localization aspect, and both properties must be satisfied
for a system to have MF modes. It has recently been shown \cite{Dumitrescu2014Majorana}
that at such a high temperature the MF signature manifested in the
tunneling experiment is diffuse over an energy range larger than the
gap itself, and as such, the issue of MF localization becomes moot
because of the participation of other states. In the second part of
our paper, we show that such broad and diffuse zero bias tunneling
conductance peak could arise from subgap non-MF states which may generically
be present in the system due to disorder. Thus, although the MF may
indeed be strongly localized in the ferromagnetic nanowire system,
tunneling experiments at temperatures (and instrumental resolution)
comparable to the gap energy cannot distinguish between MF features
and ordinary (non-MF) subgap state features. Only future experiments
carried out at much lower temperatures can therefore settle the question
of what is being observed in the experiment of Ref.~\cite{Nadj-Perge2014Observation}. At much lower temperatures, however, the strongly localized
nature of the MFs in the ferromagnetic nanowires will come into play
in a dramatic fashion, leading to a strong zero bias conductance peak
in long wires and clearly split zero bias peaks in short wires, thus
definitively establishing the existence (or not) of MFs in the ferromagnetic
wire - superconductor hybrid structure. On the other hand, if the
physics of substrate-induced MF localization length suppression is
not playing any role in the ferromagnetic adatom chains (as it is
not in semiconductor nanowires) on superconducting substrates, then
at temperatures much lower than the induced gap, the subgap zero bias
peak, if it is indeed arising from TS physics, should simply disappear
completely since the MF localization length in such an unrenormalized
situation would be much larger than the ferromagnetic adatom chain
length one can fabricate on superconducting substrates at the present
time. Thus, lowering (and improving) experimental temperature (instrumental
resolution) is the key to settling the question of whether MFs have
indeed been observed in the experiments carried out in Ref.~\cite{Nadj-Perge2014Observation}.
Since the 
state of the arts
low-temperature STM experiments are routinely carried out at $\sim 100\,{\rm mK}$
or below~\cite{Firmo2013Evidence,*Allan2013Imaging,*ZhouNatPhys2013}
(a temperature regime accessible since the early
1990s~\cite{HessPRL1990}), we urge future STM low temperature 
experiments ($<300\,{\rm mK}$)
in Fe chains on superconducting Pb substrates to settle the important
question of the existence or not of MFs in this system. 

At this stage all one can say is that while Majorana modes do not
show any basic inconsistencies with experiment, the broad features
in the experiment \cite{Nadj-Perge2014Observation} could also be
consistent with accidental fermionic non-MF subgap states. In fact,
the possibility that the observed subgap tunneling conductance structure
in the experiment of Ref.~\cite{Nadj-Perge2014Observation} arises
purely from a type of YSR bound states, rather than MF states, has
been suggested recently \cite{Sau2015Bound}. In fact, the possibility
that the observed \cite{Nadj-Perge2014Observation} broad and weak
zero bias feature arises from completely different physics \cite{Sau2015Bound,Kontos2001Inhomogeneous,*Fauchere1999Paramagnetic}
nothing to do with Majorana fermions cannot be ruled out at this stage.

\section{Majorana decay length in the Kitaev Chain}\label{sec:Kitaev}

\begin{figure}
\begin{centering}
\includegraphics[width=0.8\columnwidth]{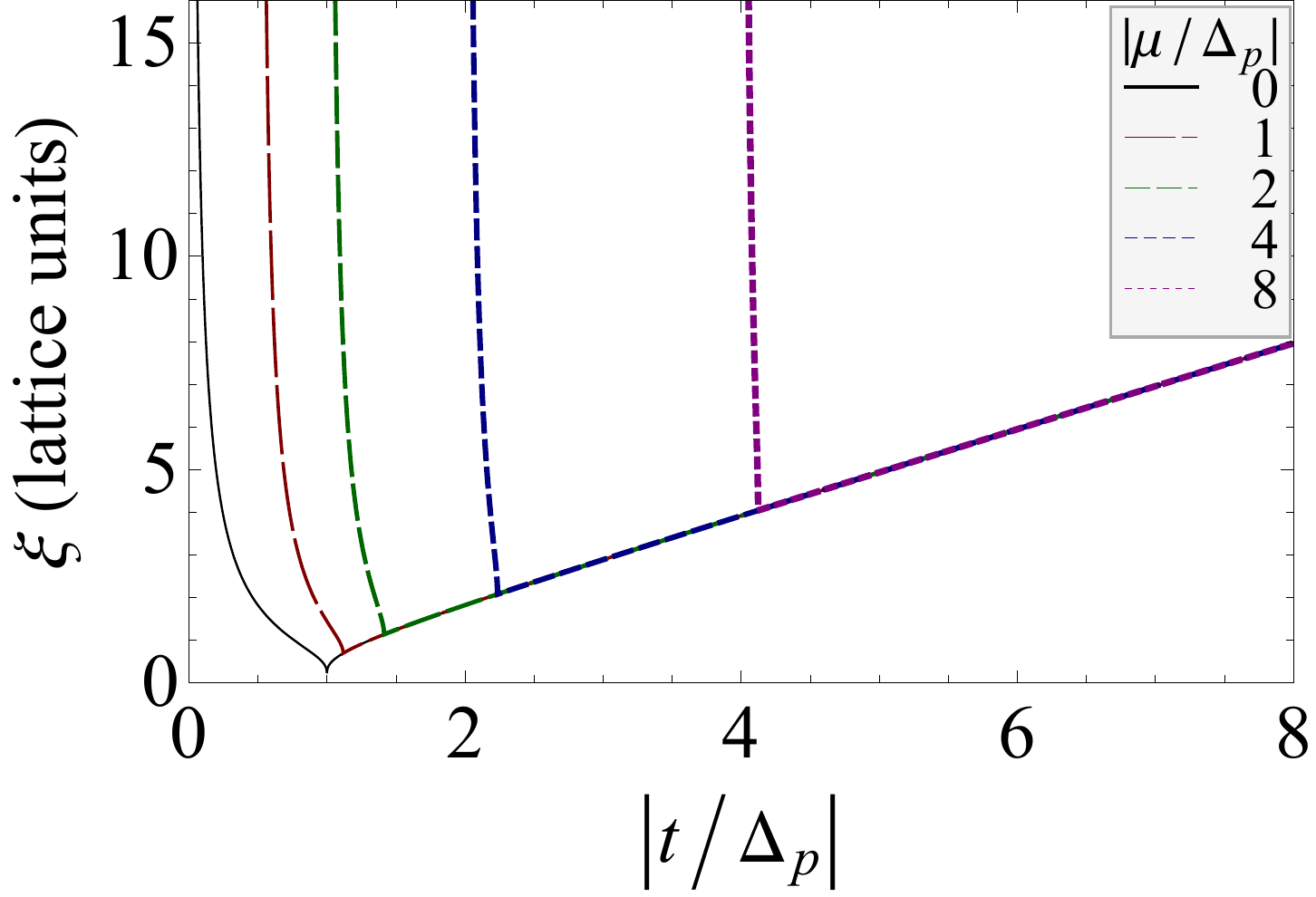} 
\par\end{centering}

\caption{\label{fig:Kloc}MF localization length $\left(\xi\right)$ given
by Eq.~(\ref{eq:xi(Kitaev)}) as a function of dimensionless hopping
strength ($|t/\Delta_{p}|$) for various values of the chemical potential
($|\mu/\Delta_{p}|$). For $|t|<|\mu|/2$ the system is in the non-topological
phase without Majorana fermions, and $\xi$ is undefined in this regime.
$\xi$ diverges at $|t|=|\mu|/2$, indicating the topological phase
transition. For $|t|\gg|\mu|$, $\xi$ is well-approximated by $|t/\Delta_{p}|$,
which is the standard coherence length formula for superconductors.}
\end{figure}

Let us first consider the prototypical and simplest model of a TS
supporting Majorana end modes, the so-called Kitaev chain. This is
a one-dimensional tight-binding model of spinless fermions with $p$-wave
pairing, described by the Hamiltonian 
\begin{align}
H_{w} & =-t\sum_{j}\left(a_{j}^{\dagger}a_{j+1}+{\rm h.c.}\right)-\mu\sum_{j}a_{j}^{\dagger}a_{j}\nonumber \\
 & +\Delta_{p}\sum_{j}\left(a_{j}a_{j+1}+{\rm h.c.}\right)\,,\label{eq:H_Kitaev}
\end{align}
 where $t$ is the hopping, $\mu$ the chemical potential, and $\Delta_{p}$
is the pairing potential. As shown by Kitaev \cite{Kitaev2001Unpaired},
this model supports unpaired Majorana modes at its boundaries for
$\left|\mu\right|<2\left|t\right|$, with a MF localization length
that is given by 
\begin{equation}
\xi=\max_{\lambda=\pm1}\left(\left|\textrm{log}\left|\frac{\lambda\sqrt{\mu^{2}-4t^{2}+4\Delta_{p}^{2}}-\mu}{2\left(t+\left|\Delta_{p}\right|\right)}\right|\right|\right)^{-1}.\label{eq:xi(Kitaev)}
\end{equation}
 We plot the localization length in Fig.~\ref{fig:Kloc} as a function
of the hopping amplitude for different values of $\mu$. Note that
$\xi$ is defined only for $\left|t\right|>\left|\mu\right|/2$, where
the system is in the topologically non-trivial regime with a Majorana
mode at each end. At $\left|t\right|=\left|\mu\right|/2$ the localization
length diverges, indicating a topological phase transition into the
topologically trivial regime at $|t|<|\mu|/2$.

There are two special limits of interest. At the special point $\mu=0$
and $|t|=|\Delta_{p}|$, the localization length vanishes and the
Majorana is localized precisely at the end site of the chain~\cite{Kitaev2001Unpaired}.
We emphasize that in this fine-tuned case the localization of the
Majorana is completely independent of the size of the energy gap,
providing a concrete example of a situation where a small gap could
in principle also be associated with a small localization length.
The other situation of interest is the physically realistic limit
$|t|\gg|\Delta_{p}|$, where the bandwidth far exceeds the superconducting
gap \cite{Sengupta2001Midgap}. Here the localization length $\xi$,
expressed to lowest order in $\Delta_{p}$, reproduces the familiar
form of a superconducting coherence length as discussed in the Introduction,
\begin{equation}
\xi=\left|\frac{t}{\Delta_{p}}\right|=\frac{v_{F}}{2E_{{\rm gap}}},\label{eq:KitaevLength}
\end{equation}
 where $E_{{\rm gap}}=|\Delta_{p}|(1-\mu^{2}/4t^{2})^{1/2}$ and $v_{F}=2|t|(1-\mu^{2}/4t^{2})^{1/2}$
are the spectral gap and Fermi velocity, respectively. Since $|t|\gg|\Delta_{p}|$,
the localization length $\xi\gg1$ and the Majorana decay length is
parametrically larger than the lattice constant (taken to be the unit
of length here). On the other hand, it is clear that if for some reasons
one can realize a Kitaev chain with $\left|t\right|\sim\left|\Delta_{p}\right|$,
as has been proposed for a quantum dot array~\cite{Sau2012Realizing},
then the MF decay length is of order a few lattice sites only, qualitatively
similar to the fine-tuned case.

\section{Substrate-Induced Renormalization of the Topological Wire}

We now turn to the physical realization \cite{Sau2010Non-Abelian,Lutchyn2010Majorana,Oreg2010Helical,Lutchyn2011Search,Brydon2015Topological,Hui2015Majorana,Dumitrescu2014Majorana,Li2014Topological,Peng2014Strong,HeimesNJP2015}
of TS in a ferromagnetic nanowire in contact with a bulk $s$-wave
superconductor. We note that the minimal model for the ferromagnetic
wire proximity-coupled to a superconductor \cite{Hui2015Majorana,Dumitrescu2014Majorana,Takei2012Microscopic},
as used in Ref.~\cite{Nadj-Perge2014Observation}, is formally the
same as the corresponding semiconductor nanowire TS system introduced
in Refs.~\cite{Sau2010Non-Abelian,Lutchyn2010Majorana,Oreg2010Helical,Lutchyn2011Search}
with the only constraint being that the spin-splitting is arising
from intrinsic exchange splitting in the ferromagnetic system whereas
it is induced as a Zeeman splitting by an externally applied magnetic
field in the semiconductor case. This has already been pointed out
by Dumitrescu \textit{et~al.} \cite{Dumitrescu2014Majorana}. The
presence of spin-orbit coupling and spin-splitting along with proximity-induced
superconductivity enables us to avoid the fermion doubling theorem,
leading to topological (effectively spinless p-wave) superconductivity
in the wire.

\subsection{Self-Energy}\label{sec:Self-Energy}

The normal state of the superconductor is characterized by strong
${\bf L}\cdot{\bf S}$ spin-orbit coupling. Furthermore, two orbitals
of different parity contribute to the states near the Fermi surface,
which we label as $s$ and $p\equiv p_{z}$, respectively. This is
not essential for our theory (for example, we could also choose three
distinct $p$-wave orbitals), but makes the following arguments more
transparent; indeed, the details of the electronic structure of the
superconductor are not important for our analysis, which relies on
symmetry considerations. Although spin itself is not a good quantum number in
the presence of spin-orbit coupling, we can label the doubly-degenerate
states by a pseudospin index $\varsigma=\pm1$, which transforms like
a spin under inversion and time-reversal. Assuming only a single band
crosses the Fermi surface, the general expression for these states
is 
\begin{equation}
\left|\bm{k},\varsigma\right\rangle =\sum_{\sigma=\uparrow,\downarrow}\left[B_{s,\varsigma\sigma}\left(\bm{k}\right)\left|s,\bm{k},\sigma\right\rangle +B_{p,\varsigma\sigma}\left(\bm{k}\right)\left|p,\bm{k},\sigma\right\rangle \right],\label{eq:pseudo}
\end{equation}
 where $B_{s,\varsigma\sigma}\left(\bm{k}\right)$ and $B_{p,\varsigma\sigma}\left(\bm{k}\right)$
are the coefficients of the $s$- and $p$-wave orbitals, respectively.
Regarding the coefficients in Eq.~(\ref{eq:pseudo}) as $2\times2$
matrices in $\varsigma-\sigma$ space, one can derive a number of
conditions. First, the normalization of the states in \eq{eq:pseudo}
requires that 
\begin{equation}
\hat{s}_{0}={B_{s}^{*}\left(\bm{k}\right)B_{s}^{T}\left(\bm{k}\right)+B_{p}^{\ast}\left(\bm{k}\right)B_{p}^{T}\left(\bm{k}\right).}\label{eq:BsBpnorm}
\end{equation}
 From inversion and time-reversal symmetries we deduce that \begin{subequations}\label{eq:BsBpcondIT}
\begin{gather}
B_{s}\left(\bm{k}\right)=B_{s}\left(-\bm{k}\right)=\hat{s}_{y}B_{s}^{*}\left(-\bm{k}\right)\hat{s}_{y}\,,\label{eq:Bsinv}\\
B_{p}\left(\bm{k}\right)=-B_{p}\left(-\bm{k}\right)=\hat{s}_{y}B_{p}^{*}\left(-\bm{k}\right)\hat{s}_{y}\,.\label{eq:Bpinv}
\end{gather}
 \end{subequations} In these equations $\hat{s}_{\mu}$ are Pauli
matrices in the $\varsigma-\sigma$ space. We additionally require
that the pseudospin index behaves like a spin under mirror reflection
in the planes perpendicular to the three Cartesian axes: \begin{subequations}\label{eq:BsBpcondM}
\begin{eqnarray}
B_{s}\left(\bm{k}\right) & = & \hat{s}_{\mu}B_{s}\left({\cal M}_{\mu}\bm{k}\right)\hat{s}_{\mu}\,,\quad\mu=x,y,z\,,\label{eq:Bsref}\\
B_{p}\left(\bm{k}\right) & = & \begin{cases}
\hat{s}_{\mu}B_{p}\left({\cal M}_{\mu}\bm{k}\right)\hat{s}_{\mu} & \mu=x,y\\
-\hat{s}_{\mu}B_{p}\left({\cal M}_{\mu}\bm{k}\right)\hat{s}_{\mu} & \mu=z
\end{cases}\label{eq:Bpref}
\end{eqnarray}
 \end{subequations}where ${\cal M}_{\mu}$ are reflection operators
for the plane perpendicular to the $\mu$-axis.

In the pseudospin basis the bulk Green function of the superconductor
is written as 
\begin{equation}
G_{SC}\left(\omega;\bm{k}\right)=\frac{\omega\hat{\tau}_{0}+\xi_{\bm{k}}\hat{\tau}_{z}+\Delta_{SC}\hat{\tau}_{x}}{\omega^{2}-\xi_{\bm{k}}^{2}-\Delta_{SC}^{2}},
\end{equation}
 where $\hat{\tau}_{\mu}$ are Pauli matrices in the particle-hole
basis, $\xi_{\bm{k}}$ is the dispersion in the superconductor, and
$\Delta_{SC}$ is the gap. The nanowire is placed on the $(001)$
surface of the superconductor. The tunneling between the two systems
is described by a superconductor-nanowire hopping term 
\begin{equation}
H_{{\rm tun}}=\sum_{n\sigma}\left[a_{n\sigma}^{\dagger}\left(t_{s}s_{n\sigma}+t_{p}p_{n\sigma}\right)+{\rm h.c.}\right],
\end{equation}
 where $a_{n\sigma}^{\dagger}$ creates an electron with spin $\sigma$
at site $n$ on the nanowire, and $s_{n\sigma}$ and $p_{n\sigma}$
are annihilation operators of the orbitals of the superconductor.
Only nearest-neighbor hopping is allowed. We remark here that a local
breaking of inversion symmetry along $z$ is required to generate
non-zero $t_{p}$ which couples the nanowire sites to the $p$-orbital
in the superconductor. Without loss of generality, we take $t_{s}$
and $t_{p}$ to be real.

The tunneling matrix $T$, which appears in the self-energy of the
nanowire {[}Eq.~(\ref{eq:tGt}){]}, necessarily includes a transformation
between the real-spin basis of the nanowire and the pseudospin basis
of the superconductor. It is written 
\begin{eqnarray}
T & = & \left(\begin{array}{cc}
t_{s}B_{s}^{T}+t_{p}B_{p}^{T} & 0\\
0 & -t_{s}B_{s}^{T}-t_{p}B_{p}^{T}
\end{array}\right).
\end{eqnarray}
 With $G_{SC}$ and $T$, the self-energy $\Sigma$ can be readily
computed as 
\begin{eqnarray}
\Sigma\left(\omega;\bm{r},\bm{r}'\right) & = & \int\frac{d^{3}k}{\left(2\pi\right)^{3}}e^{i\bm{k}\cdot\left(\bm{r}-\bm{r}'\right)}T\left(\bm{k}\right)G_{SC}\left(\omega;\bm{k}\right)T\left(\bm{k}\right)^{\dagger}\nonumber \\
 & = & \left(1-Z\right)\omega+Zt\hat{\tau}_{z}+Z\bm{\lambda}\cdot\hat{\bm{\sigma}}\hat{\tau}_{z}\nonumber \\
 &  & +Z\Delta\hat{\tau}_{x}+Z\bm{\Delta}^{(t)}\cdot\hat{\bm{\sigma}}\hat{\tau}_{x},\label{eq:selfEnergyTerms}
\end{eqnarray}
 where $Z$, $t$, $\bm{\lambda}$, $\Delta$ and $\bm{\Delta}^{(t)}$
are functions of $\omega$ and $\bm{r}-\bm{r}'$. Their full expressions
are given in the Appendix.

\subsection{Nanowire Hamiltonian}

We now consider the self-energy of the ferromagnetic nanowire in more
detail. Assuming that the nanowire lies along the $x$-axis, in the
absence of the superconductor it can be modeled by the tight-binding
Hamiltonian 
\begin{eqnarray}
H & = & t_{{\rm int}}\sum_{n\sigma}\left(a_{n\sigma}^{\dagger}a_{n+1,\sigma}+{\rm h.c.}\right)\nonumber \\
 &  & -\mu\sum_{n\sigma}a_{n\sigma}^{\dagger}a_{n\sigma}+B\sum_{n\sigma}\sigma a_{n\sigma}^{\dagger}a_{n\sigma}\,.\label{eq:barenanowireH}
\end{eqnarray}
 where $t_{{\rm int}}$ is the hopping intrinsic to the nanowire (\emph{not}
mediated by the superconductor), $\mu$ is the chemical potential,
and $B$ is the (spontaneous) ferromagnetic exchange field (written
out as an intrinsic magnetic field, rather than as an exchange splitting,
in order to maintain the explicit analogy to the semiconductor nanowire
TS platforms where $B$ is an extrinsic magnetic field).

Including the self-energy due to the proximate superconductor, the
eigenenergies of the nanowire are given by the poles of the Green
function 
\begin{equation}
G_{{\rm wire}}\left(\omega\right)=\frac{1}{\omega-H_{{\rm BdG}}-\Sigma\left(\omega\right)}=\frac{Z^{-1}}{\omega-H_{{\rm eff}}\left(\omega\right)}
\end{equation}
 where $H_{{\rm BdG}}$ is the BdG Hamiltonian of the bare nanowire~\eq{eq:barenanowireH},
and in the second equality we have rearranged terms such that the
effect of frequency renormalization is captured by $Z$, and $H_{{\rm eff}}$
contains no terms proportional to $\omega\hat{\tau}_{0}$. Explicitly,
\begin{eqnarray}
H_{{\rm eff}}\left(\omega;x_{m},x_{n}\right) & = & Z^{-1}t_{{\rm int}}\left(\delta_{m,n+1}+\delta_{m,n-1}\right)\hat{\tau}_{z}\nonumber \\
 &  & -Z^{-1}\mu\delta_{m,n}\hat{\tau}_{z}+Z^{-1}B\delta_{m,n}\hat{\sigma}_{z}\nonumber \\
 &  & +t_{m-n}\hat{\tau}_{z}+\bm{\lambda}_{m-n}\cdot\hat{\bm{\sigma}}\hat{\tau}_{z}\nonumber \\
 &  & +\Delta_{m-n}\hat{\tau}_{x}+\bm{\Delta}_{m-n}^{(t)}\cdot\hat{\bm{\sigma}}\hat{\tau}_{x}.\label{eq:Heff}
\end{eqnarray}
The subscript $m-n$ indicates that the quantities in \eq{eq:selfEnergyTerms}
are evaluated at nanowire sites with relative coordinates $\bm{r}-\bm{r}'=\left(x_{m}-x_{n}\right)\bm{e}_{x}$.

In general, the physics of the nanowire is extracted from the Green
function $G_{{\rm wire}}\left(\omega\right)$ including the frequency-dependent
self-energy. Since we are interested only in the zero-energy Majorana
mode and energy scales $\omega\ll\Delta_{0}$, however, we may take
$H_{{\rm eff}}\left(\omega=0\right)$ as our effective BdG Hamiltonian
with no loss of generality.

\subsection{Effective Kitaev models}~\label{sec:effmodel}

To make further analytical progress we need to assume specific forms
of $B_{s}$ and $B_{p}$. We take \begin{subequations}\label{Eq:BsBp}
\begin{eqnarray}
B_{s}\left(\bm{k}\right) & = & \cos\theta\hat{s}_{0},\\
B_{p}\left(\bm{k}\right) & = & \sin\theta\bm{e}_{\bm{k}}\cdot\left(\hat{s}_{y},-\hat{s}_{x},i\hat{s}_{0}\right),
\end{eqnarray}
 \end{subequations} where $\bm{e}_{\bm{k}}=\bm{k}/|\bm{k}|$. This
choice is consistent with the symmetries of the pseudospin states~\eq{eq:pseudo},
and leads to an analytically tractable result which captures the essential
physics we wish to explore. Other choices lead to qualitatively similar
results.

Using~\eq{Eq:BsBp} we calculate the full frequency-dependent forms
of $Z$, $t$, $\bm{\lambda}$, $\Delta$ and $\bm{\Delta}^{(t)}$,
which are given in the Appendix. At zero energy they take the relatively
compact forms \begin{subequations} 
\begin{eqnarray}
Z & = & 1+\frac{\Gamma}{\Delta_{SC}},\\
t_{n>0} & = & \frac{\Gamma}{Z}\frac{\cos n\tilde{a}}{n\tilde{a}}e^{-n/\tilde{\xi}},\\
\bm{\lambda}_{n>0}\cdot\hat{\bm{\sigma}} & = & -i\hat{\sigma}_{y}g\sin2\theta\frac{\Gamma}{Z}\frac{\cos n\tilde{a}+n\tilde{a}\sin n\tilde{a}}{n^{2}\tilde{a}^{2}}e^{-n/\tilde{\xi}}\\
\Delta_{n} & = & \frac{\Gamma}{Z}\frac{\sin n\tilde{a}}{n\tilde{a}}e^{-n/\tilde{\xi}}\\
\bm{\Delta}_{n>0}^{(t)}\cdot\hat{\bm{\sigma}} & = & i\hat{\sigma}_{y}g\sin2\theta\frac{\Gamma}{Z}\frac{\sin n\tilde{a}-n\tilde{a}\cos n\tilde{a}}{n^{2}\tilde{a}^{2}}e^{-n/\tilde{\xi}}\,,
\end{eqnarray}
\label{eq:parameters} \end{subequations}where we have introduced
dimensionless variables 
\begin{equation}
\Gamma=\pi\nu\left(t_{s}^{2}\cos^{2}\theta+t_{p}^{2}\sin^{2}\theta\right),\label{eq:Gamma}
\end{equation}
 $g=\pi\nu t_{s}t_{p}$, $\tilde{a}=k_{F}a_{{\rm lat}}$, and $\tilde{\xi}=\xi/a_{{\rm lat}}$,
in which $\nu$ and $\xi$ are respectively the Fermi-level density
of states and the coherence length of the superconductor, and $a_{{\rm lat}}$
is the lattice constant of the tight-binding model of the nanowire
{[}Eq.~(\ref{eq:barenanowireH}){]}.

In the limit of large exchange field
$\left(B\gg\left|\bm{\lambda}_{n}\right|,\left|\bm{\Delta}_{n}\right|\right)$,
the effects of the spin-orbit coupling and $s$-wave pairing terms 
are suppressed. As shown in~Appendix~\ref{appendix:Heff}, the Hamiltonian
(\ref{eq:Heff}) thus reduces to two copies of the Kitaev 
model, albeit with long-range hopping. To make connections with Sec.~\ref{sec:Kitaev},
we first ignore the long-ranged part of the self-energy which is beyond
nearest neighbors, yielding (second quantized) effective Hamiltonians
for spin-up $\left(+\right)$ and spin-down $\left(-\right)$ species
\begin{eqnarray}
H_{0}^{(\pm)} & = & -\left(Z^{-1}t_{{\rm int}}+t_{1}\right)\sum_{j}\left(a_{j}^{\dagger}a_{j+1}+{\rm h.c.}\right)\nonumber \\
 &  & -Z^{-1}\left(\mu\mp B\right)\sum_{j}a_{j}^{\dagger}a_{j}\nonumber \\
 &  & +\Delta_{1}^{(t)}\sum_{j}\left(a_{j}a_{j+1}+{\rm h.c.}\right).\label{eq:Heffspin}
\end{eqnarray}
 The induced hopping integral and pairing potential are given by 
\begin{eqnarray}
t_{1} & = & \frac{\Gamma}{Z}\frac{\cos\tilde{a}}{\tilde{a}}e^{-1/\tilde{\xi}}\,,\\
\Delta_{1}^{(t)} & = & g\sin2\theta\frac{\Gamma}{Z}\frac{\sin\tilde{a}-\tilde{a}\cos\tilde{a}}{\tilde{a}^{2}}e^{-1/\tilde{\xi}}\,.
\end{eqnarray}
 From Eq.~(\ref{eq:KitaevLength}), the localization length for the
Majorana zero modes in these Hamiltonians (valid in their topological
phase) is thus 
\begin{equation}
\xi_{{\rm wire}}=\frac{Z^{-1}t_{{\rm int}}+t_{1}}{\Delta_{1}^{(t)}}=\frac{\tilde{v}_{F}^{(\pm)}}{2E_{{\rm gap}}^{(\pm)}}\label{eq:HeffLength}
\end{equation}
 where the renormalized Fermi velocity and excitation gap are 
\begin{eqnarray}
\tilde{v}_{F}^{(\pm)} & = & 2\left(Z^{-1}t_{{\rm int}}+t_{1}\right)\sqrt{1-\frac{\left(\mu\mp B\right)^{2}}{4\left(t_{{\rm int}}+Zt_{1}\right)^{2}}}\,,\\
E_{{\rm gap}}^{(\pm)} & = & \Delta_{1}^{(t)}\sqrt{1-\frac{\left(\mu\mp B\right)^{2}}{4\left(t_{{\rm int}}+Zt_{1}\right)^{2}}}\,.
\end{eqnarray}
 Note that the localization length is the same for the spin-up and
-down sectors. From Eq.~(\ref{eq:HeffLength}) we observe that if
one ignores the renormalization of the Fermi velocity and uses instead
its intrinsic value, $v_{F}^{(\pm)}=2t_{{\rm int}}[1-\left(\mu\mp B\right)^{2}/4t_{{\rm int}}^{2}]^{1/2}$,
to estimate the localization length as $\xi=v_{F}^{(\pm)}/2E_{{\rm gap}}^{(\pm)}$,
the result would \emph{overestimate} the true value by a factor of
$v_{F}^{(\pm)}/\tilde{v}_{F}^{(\pm)}>1$. If the coupling between
the wire and the superconductor is weak (i.e. $\Gamma\ll\Delta_{SC},t_{{\rm int}}$),
the velocity is only weakly renormalized and $\tilde{v}_{F}^{(\pm)}\approx v_{F}^{(\pm)}$.
However, when $\Gamma$ is comparable to $\Delta_{SC}$ or even $t_{{\rm int}}$,
the discrepancy between the renormalized and the bare Fermi velocity
is huge. For large enough $\Gamma$ and hence $Z$, the coherence
length could be close to zero even though the induced triplet gap
is small. Whether or not this strong velocity renormalization, leading
to sharply-localized MFs in the TS nanowire, is present in the experiment
of Ref.~\cite{Nadj-Perge2014Observation} can only be determined
empirically since the microscopic details about $\Gamma$ are simply
not known in the experimental system. What is clear, however, is that
there is a well-defined physical mechanism, namely, a very strong
tunnel-coupling between the superconductor and the nanowire, which
would lead to a strong renormalization of the effective Fermi velocity
and a concomitant suppression of the MF localization length in the
nanowire even if the induced topological gap is small. We note that
the existence of the strong renormalization effect has already been
invoked for the Fe/Pb system by Peng \emph{et al.} using a helical
magnetic chain model for the nanowire~\cite{Peng2014Strong}.

\subsection{Effects of non-local hopping and pairing}

As we mentioned in the introduction, the substrate-induced enhancement
of the Majorana localization length is accompanied by a power-law
decay of the MFs \cite{Pientka2013Topological}. This power-law decay
of the MFs, if large, limits the validity of the enhanced exponential
localization. To understand and estimate this effect we write the
Hamiltonian in the large tunneling limit as 
\begin{equation}
H=H_{0}+\delta H
\end{equation}
 where $H_{0}$ is given in Eq.~(\ref{eq:Heffspin}) and $\delta H$
contains the hopping and pairing terms in Eq.~(\ref{eq:Heff}) involving
sites separated by two or more lattice spacings. Let $\psi_{0}$ denote
the zero-energy Majorana mode that is localized at the end of the
wire with a localization length given by Eq.~(\ref{eq:HeffLength}).
With the non-local perturbation $\delta H^{(\pm)}$ the state acquires
a correction $\psi_{0}\rightarrow\tilde{\psi}_{0}=\psi_{0}+\delta\psi_{0}$,
where 
\begin{equation}
\delta\psi_{0}=-\frac{1}{H_{0}}{\cal P}\delta H\left|\psi_{0}\right\rangle \,,
\end{equation}
where ${\cal P}=1-\left|\psi_{0}\right\rangle \left\langle \psi_{0}\right|$.
We can now qualitatively see the localization behavior of
  $\tilde{\psi}_{0}$ including the long-range self-energy
  correction. We begin by re-writing the equation for $\delta\psi_0$
  in real space  with coordinate $x$ as
\begin{equation}
\delta\psi_{0}(x)=-\int dx' dx_1 \tilde{G}_0(x,x')\delta H(x',x_1)\psi_{0}(x_1) \,\label{eqdpsi0},
\end{equation} 
where $\tilde{G}_0(x,x')\approx H_0^{-1}\mathcal{P}$ is the projected
Green function  corresponding to the Hamiltonian $H_0$ at zero
energy. For a gapped system the Green function  
$\tilde{G}_0(x,x')\propto e^{-|x-x'|/\xi_{{\rm wire}}}$ and the unperturbed 
wavefunction 
$\psi_{0}\sim e^{-x/\xi_{{\rm wire}}}$ are both  localized on a length
scale $\xi_{{\rm wire}}$.
On the other hand, the perturbation $\delta H$ has a power-law tail so that $\delta H(x',x_1)\propto 
e^{-|x'-x_1|/\xi}/(k_{F}|x'-x_1|)$ where $\xi$ is the coherence length
of the bulk superconductor. Assuming that $\xi_{{\rm wire}}\ll \xi$,
we see that the integral in Eq.~\ref{eqdpsi0} 
is dominated by $|x_1|,|x-x'|\lesssim \xi_{{\rm wire}}$ so that for $x\gg \xi_{{\rm wire}}$
the perturbed wave-function is written as 
\begin{equation}
\tilde{\psi}_{0}=\psi_0+\delta\psi_0\sim\alpha e^{-x/\xi_{{\rm wire}}}+\beta\frac{e^{-x/\xi}}{k_{F}x},\label{eq:perturbedpsi}
\end{equation}
where $\alpha$ is a normalization constant and $\beta$ is a parameter
determined from the perturbation theory. Strictly speaking, the localization
length of $\tilde{\psi}_{0}$ is $\max\left\{ \xi_{{\rm wire}},\xi\right\} $.
However, the factor of $k_F$ in the denominator of the second term in
Eq.~(\ref{eq:perturbedpsi}) means that this component of the
wavefunction is effectively localized on a scale of $k_{F}^{-1} \ll
\xi$. This recalls YSR bound states around magnetic
impurities in conventional
superconductor~\cite{Yu1965Bound,*Shiba1968Classical,*Rusinov1968BSuperconcductivity},
where the bound state wavefunction has a similar spatial profile in
the superconductor. Indeed, experiments show that such bound states
are localized on a scale much smaller than
$\xi$~\cite{Yazdani1997Probing}. 

The implication of~\eq{eq:perturbedpsi} is that since the dominant
localization length of the non-local part is small,  the non-local
term $\delta H$ can be safely ignored: the experimentally measured
localization length will still be $\xi_{{\rm wire}}$ even with
longer-range hopping in Eq.~\ref{eq:Heff}.
We note, however, that independent of whether $\alpha\gg\beta$ (the
perturbative regime) or
$\beta\gg\alpha$ (the YSR regime) in
Eq.~(\ref{eq:perturbedpsi}), the resultant 
Majorana wavefunction is strongly localized at the wire end ($x=0$)
with a localization length of $\xi_{\rm wire}$
or $k^{-1}_{F}$, both of which are much smaller than the bare MF localization
length $\xi$ expected without the substrate renormalization effect
(provided, of course, one is in the strong tunnel coupling regime).
In Ref.~\cite{Dumitrescu2014Majorana}, 
Dumitrescu \textit{et~al.} recently took into account the second
term in Eq.~(\ref{eq:perturbedpsi}) as causing the suppressed MF
localization in ferromagnetic chain TS systems whereas Peng \textit{et~al.}
\cite{Peng2014Strong} mostly considered the first term in discussing
MF localization in helical magnetic chains. In principle, both terms
could be important, but their qualitative effects are similar, both
leading to a strongly suppressed MF localization in the nanowire in
the strong tunnel-coupled regime. Importantly, the power law
decay of the second term in Eq.~(\ref{eq:perturbedpsi})
 may have negative
implications for the non-Abelian braiding
experiments~\cite{Zyuzin2013Correlations}, severely limiting the usefulness of the resultant MFs in carrying out topological
quantum computation although for practical purposes the MFs appear
strongly spatially localized at the wire ends.

\subsection{Relating quasi-1D models to 1D models\label{sub:Dependence-of-Gamma}}

We have established above that as long as the nanowire is strongly
tunnel-coupled to the superconductor (so that the condition $\Gamma>t_{{\rm int}},\Delta_{{\rm SC}}$
applies), the MF localization length would be strongly suppressed
compared with the standard bare coherence length formula due to the
Fermi velocity renormalization caused by the substrate. This renormalization
effect appears to be independent of the nature of the nanowire and,
therefore, should affect both ferromagnetic nanowires and semiconductor
nanowires equally (as long as the tunnel coupling defined by Eq.~(\ref{eq:Gamma})
is large). We now show that this is not the case, and there is good
reason to believe that the ferromagnetic chain system of Ref.~\cite{Nadj-Perge2014Observation}
is much more strongly renormalized by the substrate than the
semiconductor nanowire systems \cite{Mourik2012Signatures,Deng2012Anomalous,Rokhinson2012fractional,Das2012Zero-bias,Churchill2013Superconductor-nanowire,Finck2013Anomalous}.

While it is common to assume a strictly 1D limit in modeling both the
semiconductor nanowire and ferromagnetic chain systems, a more
realistic model would treat them as quasi-1D and as a result
the parameters $t_s$ and 
$t_p$ in the 1D model [for example, defining $\Gamma$
  in~\eq{eq:Gamma}] are really effective parameters that have a strong
dependence on the radius of the nanowire in a quasi-1D
geometry. Since  we are interested in understanding the scaling behavior
with nanowire radius, we will assume a simple model of a
three-dimensional (3D) cylindrical lattice
nanowire (with the wire transverse cross-sectional width being much
smaller than the wire length). The 3D (i.e. quasi-1D)
wavefunctions and the strictly 1D wavefunctions $\psi^{{\rm wire,1D}}(x)$
are related by a transverse wavefunction factor as 
\begin{eqnarray}
\psi^{{\rm wire,3D}}_{m,j}\left(\rho,x,\phi\right) &=
&\frac{a}{\sqrt{\pi}(R+a)J^{\prime}_m(k_{m,j})}J_{m}\left(k_{m,j}\frac{\rho}{R+a}\right)
\notag \\
&& \times e^{im\phi}\psi^{{\rm wire,1D}}\left(x\right)\label{eq:tsest}
\end{eqnarray}
where $k_{m,j}$ is the $j^{\text{th}}$ zero of the $m^{\text{th}}$
Bessel function so that the wavefunction 
satisfies $\psi^{{\rm wire,3D}}_{m,j}(R+a,x,\phi)=0$ and $a$ is the lattice
constant of the wire. In deriving this expression we have used the
continuum limit $k_{m,j} a/R\ll 1$ so that the {\it lattice}
wavefunction can be 
approximated as $a$ times the {\it continuum} wavefunction of
a cylinder. Requiring that the lattice wavefunction vanishes at the
edge of the cylinder corresponds to the boundary condition that the
continuum wavefunction vanishes a distance $a$ outside the wire.

The 1D hopping matrix elements $t_{s,p}$ enter the
formalism in Sec.~\ref{sec:effmodel}  
through the parameter $\Gamma$ defined in Eq.~\ref{eq:Gamma}. To
simplify our analysis we split
$\Gamma=\Gamma_{s}\cos{\theta}^{2}+\Gamma_{p}\sin{\theta}^{2}$ 
where $\Gamma_{s}=\pi\nu t_{s}^{2}$ and $\Gamma_{p}=\pi\nu t_{p}^{2}$.
The matrix elements $t_{s,p}$ are proportional to the absolute
value square of the
3D tunnelings $\tilde{t}_{s,p}$ and the 
quasi-1D wavefunction $\psi^{{\rm
    wire,3D}}_{m,j}\left(\rho,x,\phi\right)$ at the surface of the wire.
In addition, for the purpose of our estimate, we will make a simplifying
assumption that the Green function of the substrate (which determines
the density of states $\nu=\textrm{Im}\{G\}$) is local,
i.e. $k_Fa\approx 1$. With these   
assumptions (which may be relaxed without qualitatively changing  the
results), the generalized form for $\Gamma_{s,p}$ is
written 
\begin{eqnarray}
\Gamma_{s,p} & = & \pi \nu \tilde{t}_{s,p}^{2}\sum_{\text{surf}}|\psi^{{\rm wire,3D}}_{m,j}\left(R,x,\phi\right)|^{2}\notag \\
&\approx & \pi \nu \tilde{t}_{s,p}^{2} \frac{R}{a}\int d\phi dx|\psi^{{\rm wire,3D}}_{m,j}\left(R,x,\phi\right)|^{2},
\end{eqnarray}
where the sum in the first line is over all lattice sites at the
surface of the wire. For $R\gg a$ the wavefunction at the surface is
approximately given by
\begin{equation}
\psi^{{\rm wire,3D}}_{m,j}\left(R,x,\phi\right) \approx -\frac{a^2
  k_{m,j}}{\sqrt{\pi}R^2}e^{im\phi}\psi^{{\rm wire,1D}}(x)\,,
\end{equation}
and so we find
\begin{equation}
\Gamma_{s,p} = \nu \tilde{t}_{s,p}^{2} \frac{a^3 k_{m,j}^2}{ R^3}\,.
\end{equation}
The hopping $\tilde{t}_{s,p}$ can be parametrized by a dimensionless parameter
$\zeta_{s,p}$ and written as $\tilde{t}_{s,p}=\zeta_{s,p}E_F$. Since
the hopping between the wire and the SC is expected to be of the same
order of magnitude as the bare hopping in the substrate, we expect
$\zeta_{s,p}$ to be a parameter of order $1$. while 
the density of states is $\nu = k^3_Fa^3/4\pi^2E_F$.
 Using this parametrization we estimate $\Gamma_{s,p}$ as 
\begin{equation}
\Gamma_{s,p}\approx\kappa\left(\frac{a}{R}\right)^{3}E_{F}\,,\label{eq:GammaDependence}
\end{equation}
where $\kappa = \zeta_{s,p}^2k_{m,j}^2(k_Fa)^3/4\pi^2 $.

We are now in a position to compare $\Gamma$ for the
semiconductor nanowire and the ferromagnetic chain. Qualitatively
speaking, $R$ is much smaller for a ferromagnetic chain of
atoms (e.g. Ref.~\cite{Nadj-Perge2014Observation}) than for the
semiconductor nanowire 
(e.g. Ref.~\cite{Mourik2012Signatures}), and so $\Gamma$ is
accordingly expected
to be much larger in the former compared to the
latter due to the dependence $\Gamma \propto R^{-3}$. Quantitatively,
assuming $R\sim a$  
is of order $0.5\,{\rm nm}$ for the ferromagnetic Fe chain
in Ref.~\cite{Nadj-Perge2014Observation}, we estimate $\Gamma\sim
E_{F}\sim1{\rm eV}$ (assuming
$\kappa\sim1$). On the other hand, for the 
semiconductor nanowire with much larger mode confinement radius
$R\sim20{\rm nm}$, the same values for the other parameters yields a
self-energy parameter $\Gamma$ on the order $0.01{\rm meV}$. This huge
difference in $\Gamma$ between the semiconductor 
nanowires used in \cite{Mourik2012Signatures} and the ferromagnetic
chains used in \cite{Nadj-Perge2014Observation} gives a natural
explanation for why the
MF might be strongly localized (delocalized) in the ferromagnetic
(semiconductor) nanowires even if both systems manifest the same
induced superconducting gap ($\sim100\,{\rm \mu eV}$). This difference
arises, keeping all the other parameters similar, from
the difference in the transverse quantization size in the two
systems. The wire radius ratio of roughly a factor of 40 between them
can in principle lead to a localization length difference by as large
as a factor of $40^{3}=64000$! In reality, this is 
an overestimate, since the bare Fermi velocity in the semiconductor is
typically a factor of 100 or so smaller than that in the ferromagnetic
metallic chain and also the tunneling factor $\zeta_{s,p}$ in the semiconductor 
can easily be smaller by an order of magnitude. The combination of 
these factors may reduce the factor from $64000$ to $64000/100\sim640$
difference in 
the MF localization length between the semiconductor nanowire \cite{Mourik2012Signatures,Deng2012Anomalous,Rokhinson2012fractional,Das2012Zero-bias,Churchill2013Superconductor-nanowire,Finck2013Anomalous}
and the ferromagnetic wire \cite{Nadj-Perge2014Observation} systems
even if both have exactly the same induced superconducting
gap ($\sim0.1\,{\rm meV}$). This factor of $\sim500$ difference
is in quantitative agreement with the conclusion of Ref. 21 where
the MF localization length is inferred to be $<1\,{\rm nm}$ whereas
in the semiconductor nanowire case the MF localization length is the
same as the bare coherence length in the nanowire ($\sim100\,{\rm nm}$).
Thus, the difference between MF localization in the two systems arises
entirely from the difference in the nanowire transverse confinement
radius in semiconductors versus metals.

In the next section we show that finite temperature effects in the
experiment of \cite{Nadj-Perge2014Observation} would make the Majorana
zero-mode signature weak and diffuse in the tunneling conductance
measurement (in spite of strong MF localization) because of strong
thermal broadening of both the MF mode and above-gap fermionic
excitations in the system. Thus, the high-temperature MF signature
(at temperatures comparable to the gap energy) is qualitatively similar
in the tunneling spectroscopy as that of any generic non-MF subgap
excitation even when these fermionic subgap excitations are not at
zero energy. Only experiments at temperatures much lower than the
induced gap energy can distinguish MF versus ordinary fermionic subgap
states in the tunneling spectra independent of the localization properties
of the Majorana zero modes. At temperatures much lower than the gap
energy, however, the strongly suppressed MF localization length in
the ferromagnetic nanowire becomes an extremely important physical
quantity since the very short MF localization length may now allow
well-defined (rather than strongly hybridized) zero-energy MF modes
to exist in rather short magnetic chains used in Ref.~\cite{Nadj-Perge2014Observation},
which would not be possible if the MF localization length is given
by the bare formula. Whether this physics is operational or not can
only be determined empirically by carrying out STM measurements at
temperatures much lower than the gap energy.

\section{Impurity-Induced Subgap States}

Split Majoranas that may result from a wire of length comparable to
the localization length are essentially identical to conventional
Andreev states that are of non-topological origin. In this section
we examine tight-binding models for the tunneling conductance from
an STM tip into either a $p$-wave or an $s$-wave nanowire with subgap
states of nontopological origin. In the $p$-wave case, we induce
subgap Andreev states by including non-magnetic impurities. In the
$s$-wave case, on the other hand, YSR \cite{Yu1965Bound,*Shiba1968Classical,*Rusinov1968BSuperconcductivity}
subgap states are created by magnetic impurities. The calculated finite
temperature tunneling conductance results (Fig.~\ref{fig:subgap})
due to these fermionic subgap states are compared with the corresponding
MF-induced tunneling spectra in a TS (Fig.~\ref{fig:Majorana}).
We demonstrate that, in general, the two sets of results are almost
impossible to distinguish unless the experimental temperature and
energy resolution are much smaller than the superconducting gap energy.
We mention here that the temperature dependence of the MF tunneling
conductance spectra for the ferromagnetic nanowire system has already
been calculated in great details by Dumitrescu \textit{et~al.} \cite{Dumitrescu2014Majorana},
who have established that high-temperature ZBCP is spectrally spread
over the whole energy gap as a very weak and very broad feature making
the interpretation of the data in Ref.~\cite{Nadj-Perge2014Observation}
problematic. Our results for the temperature dependence of MF induced
ZBCP in the ferromagnetic nanowires agree completely with the results
presented in Ref.~\cite{Dumitrescu2014Majorana}, but what is new
in our current work is showing that non-MF subgap states may also
lead to tunneling conductance features which are indistinguishable
from the corresponding MF features in high temperature experiments.
This comparison between MF versus non-MF conductance features is the
new ingredient in our results.

\subsection{Model for a $p$-wave nanowire\label{sub:p-model}}

The Hamiltonian describing tunneling into the $p$-wave nanowire is
\begin{subequations}\label{eq:Hp} 
\begin{align}
H & =H_{w}+H_{{\rm lead}}+H_{{\rm tun}}\\
H_{w} & =-t\sum_{j}\left(a_{j}^{\dagger}a_{j+1}+{\rm h.c.}\right)-\sum_{j}\mu_{j}a_{j}^{\dagger}a_{j}\nonumber \\
 & +\Delta_{p}\sum_{j}\left(a_{j}a_{j+1}+{\rm h.c.}\right)\label{eq:Hw}\\
H_{{\rm tip}} & =\sum_{k}\epsilon_{k}c_{k}^{\dagger}c_{k}\\
H_{{\rm tun}} & =t_{L}c_{k}^{\dagger}a_{1}+{\rm h.c.}
\end{align}
 \end{subequations} where $H_{w}$, $H_{{\rm tip}}$ and $H_{{\rm tun}}$
are respectively the Hamiltonians for the nanowire, the STM tip, and
the tunneling from the tip to the nearest site on the nanowire (designated
as site $1$). The annihilation operator for site $j$ in the nanowire
is $a_{j}$, while the annihilation operator for state $k$ in the
STM tip is $c_{k}$. The $p$-wave spectral gap of the nanowire is
$\Delta=\Delta_{p}(1-\mu^{2}/4t^{2})^{1/2}$. Non-magnetic impurities
are introduced by giving the chemical potential a site-dependence.
We ignore spin as we are only interested in non-magnetic impurities.

The zero-temperature conductance is computed from the Green function
${\cal G}\left(x,x',\omega\right)$ of the nanowire by \cite{Lobos2014Tunneling}
\begin{equation}
G_{T=0}\left(\omega\right)=\frac{e^{2}}{h}\left[2\pi\gamma\rho_{1}\left(\omega\right)-\gamma^{2}\left|g_{11}^{2}\left(\omega\right)\right|+\gamma^{2}\left|f_{11}^{2}\left(\omega\right)\right|\right]
\end{equation}
 where $g$ and $f$ are respectively the normal and anomalous parts
of ${\cal G}$, and the subscripts {}``11'' indicate that both position
arguments of the Green function are at the nanowire site in contact
with the STM tip. $\gamma=2\pi\nu t_{L}^{2}$ is the broadening due
to the STM tip and $\rho_{1}=-(1/\pi){\rm Im}g_{11}$ is the local
density of states at the point of contact with the tip. Numerically,
the Green function is given by 
\begin{equation}
{\cal G}\left(x,x',\omega\right)=\left[\omega+i\delta-H_{w}\left(x,x'\right)+i\frac{1}{2}\Gamma\left(x,x'\right)\right]^{-1}
\end{equation}
 where $\Gamma\left(x,x'\right)=\gamma\delta_{x,1}\delta_{x',1}$
is the broadening induced by the contact to the STM tip. The broadening
term $i\delta$ mimics the energy resolution of the setup, with a
lower $\delta$ implying a higher resolution. We note that the experiment
of Ref.~\cite{Nadj-Perge2014Observation} has a very poor energy
resolution, comparable to the induced energy gap in the nanowire,
making the broadening parameter $\delta$ an important aspect of the
experimental analysis.

The finite-temperature conductance can be obtained from $G_{T=0}$
using 
\begin{eqnarray}
G_{T}\left(\omega\right) & = & \int d\varepsilon G_{T=0}\left(\varepsilon\right)\partial_{\epsilon}n_{F}\left(\omega-\epsilon\right)\nonumber \\
 & = & \frac{1}{4T}\int d\varepsilon G_{T=0}\left(\varepsilon\right){\rm sech}^{2}\left(\frac{\omega-\epsilon}{2T}\right)
\end{eqnarray}
 where $n_{F}\left(\omega\right)$ is the Fermi distribution function.
The thermal broadening of the zero-temperature conductance is also
important to understanding the high-temperature STM experiments carried
out in Ref.~\cite{Nadj-Perge2014Observation} as already emphasized
in Ref.~\cite{Dumitrescu2014Majorana}.

\subsection{Model for an $s$-wave nanowire\label{sub:s-model}}

The Hamiltonian for tunneling into the $s$-wave nanowire is\begin{subequations}\label{eq:Hs}
\begin{align}
H & =H_{w}+H_{{\rm lead}}+H_{{\rm tun}}\\
H_{w} & =-t\sum_{j\sigma}\left(a_{j\sigma}^{\dagger}a_{j+1,\sigma}+{\rm h.c.}\right)-\mu\sum_{j\sigma}a_{j\sigma}^{\dagger}a_{j\sigma}\nonumber \\
 & +\sum_{j\sigma}B_{j}\sigma a^\dagger_{j\sigma}a_{j\sigma}+\Delta_{s}\sum_{j}\left(a_{j\uparrow}a_{j\downarrow}+{\rm h.c.}\right)\label{eq:Hw-1}\\
H_{{\rm tip}} & =\sum_{k\sigma}\epsilon_{k}c_{k\sigma}^{\dagger}c_{k\sigma}\\
H_{{\rm tun}} & =t_{L}\sum_{\sigma}c_{k\sigma}^{\dagger}a_{1\sigma}+{\rm h.c.}
\end{align}
 \end{subequations}We again adopt the convention that the nanowire
site in contact with the STM tip is denoted as site $j=1$. The definition
of the annihilation operators is generalized to include the spin degrees
of freedom, which must be accounted for in this case. In contrast
to the $p$-wave nanowire, here we take a uniform chemical potential
$\mu$, but allow for the possibility of magnetic impurities through
the site-dependent Zeeman field $B_{j}$. The spectral gap is $\Delta=\Delta_{s}$.

The analysis of the tunneling proceeds similarly to that for the $p$-wave
nanowire above. The expression for the zero-temperature conductance
is, however, slightly modified to include the spin degrees of freedom
\begin{align}
G_{T=0}\left(\omega\right) & =\frac{e^{2}}{h}\left[2\pi\gamma\sum_{\sigma}\rho_{1\sigma}\left(\omega\right)\right.\nonumber \\
 & \left.-\gamma^{2}\sum_{\sigma\sigma'}\left(\left|g_{1\sigma,1\sigma'}^{2}\left(\omega\right)\right|-\left|f_{1\sigma,1\sigma'}^{2}\left(\omega\right)\right|\right)\right]
\end{align}
 where $g$ and $f$ are the normal and anomalous Green functions,
respectively, and the subscripts indicates the site and spin indices.

\subsection{Effects of high temperature and low resolution on the tunneling conductance}

\begin{figure*}
\begin{centering}
\includegraphics[width=1\textwidth]{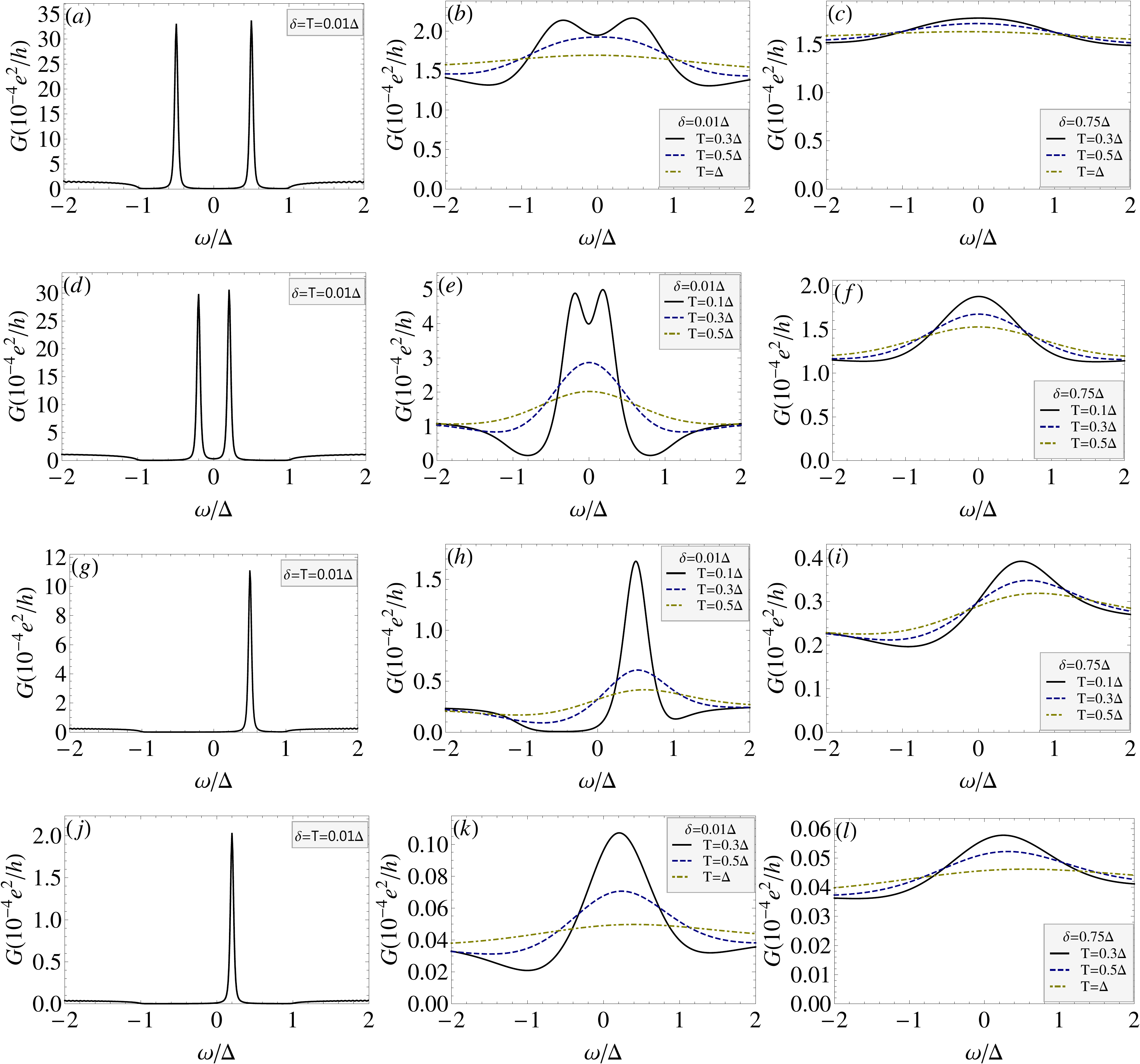} 
\par\end{centering}

\caption{(a-c) The conductance measured by placing the STM tip at a site with
a magnetic impurity in an infinite $s$-wave nanowire with $t=2\mu=10\Delta$
and $\gamma=0.001\Delta$. The magnetic impurity induces a YSR state
with subgap energy $0.5\Delta$. We show the conductance at high resolution
and low temperature {[}panel (a){]}, high resolution and high temperature
{[}panel (b){]}, and low resolution {[}panel (c){]}. (d-f) Similar
to Panels (a-c) but with smaller subgap energy $0.2\Delta$. (g-i)
Similar to Panels (a-c) but at the site of a nonmagnetic impurity
in an infinite $p$-wave nanowire with $t=2\mu=10\Delta$. The impurity
induces an Andreev state with energy $0.5\Delta$. (j-l) Similar to
Panels (g-i) but with smaller subgap energy $0.2\Delta$. \label{fig:subgap}}
\end{figure*}


Fig.~\ref{fig:subgap} shows the differential conductance obtained
by tunneling into nontopological subgap states in an infinite nanowire,
induced by either magnetic impurities (in the $s$-wave nanowire)
or non-magnetic impurities (in the $p$-wave nanowire). At high resolution
and low temperature (Fig.~\ref{fig:subgap}a,d,g,j) the exact non-zero
energy of the subgap state can be extracted from the tunneling spectra.
With either high temperatures (Fig.~\ref{fig:subgap}b,e,h,k) or
low resolutions (Fig.~\ref{fig:subgap}c,f,i,l), however, the peaks
broaden and could appear to arise from a broadened zero-energy peak.
In fact, Fig.~\ref{fig:subgap} shows that with high temperature
or low resolution (or both), the tunneling conductance results in
the nanowires typically manifest broad features consistent with a
zero-energy peak as long as the subgap states are located at $|E|\lesssim0.5\Delta$.
Thus, the observation of broad zero-bias conductance features should
not be associated as evidence for the existence of precisely zero-energy
MFs since this is also consistent with fermionic subgap states such
as ordinary YSR or Andreev bound states. Indeed, the possibility
that the experiment of Ref.~\cite{Nadj-Perge2014Observation} is
actually observing a YSR state feature instead of a MF state has
recently been suggested in the literature \cite{Sau2015Bound}.

\begin{figure}
\begin{centering}
\includegraphics[width=1\columnwidth]{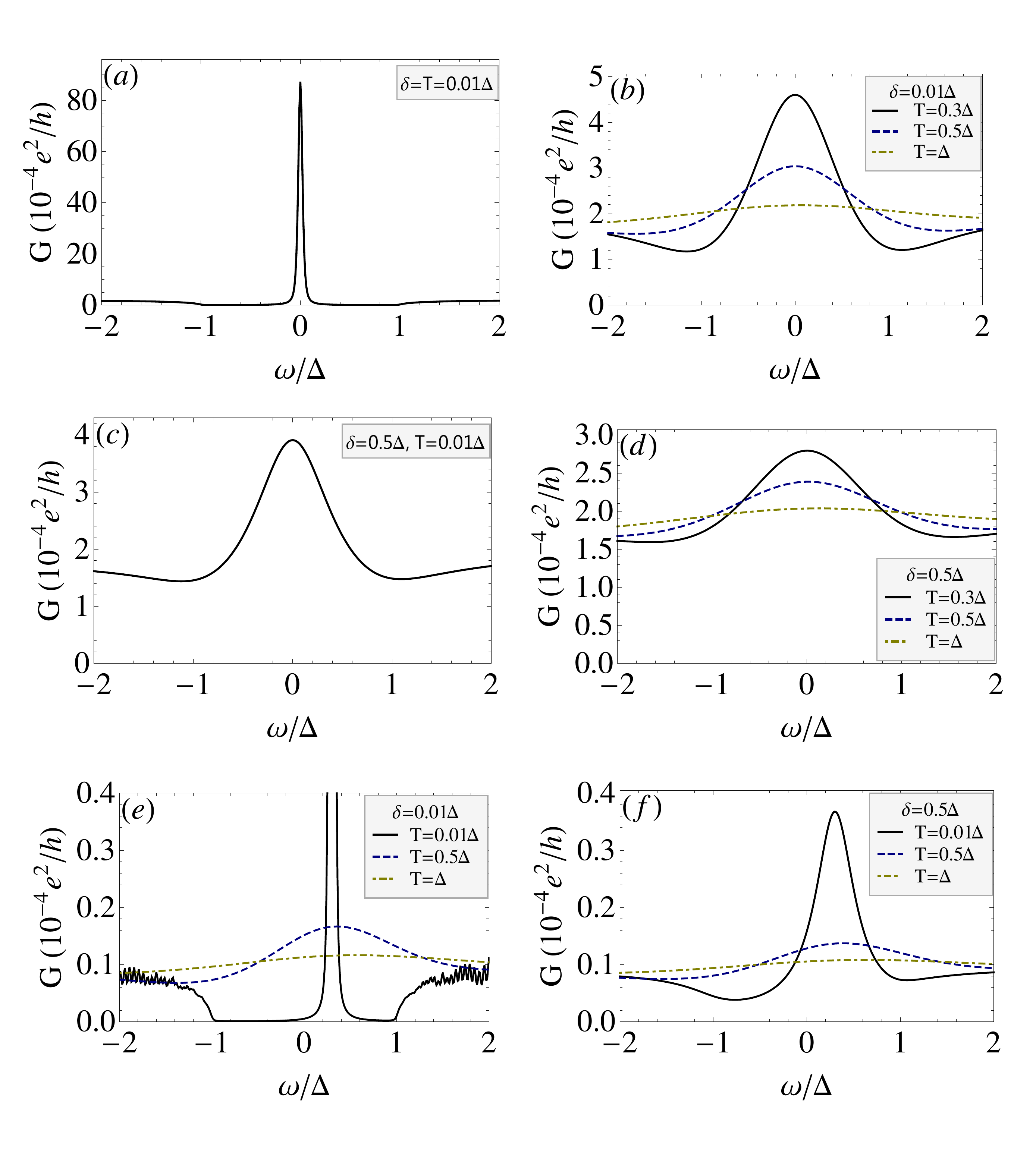} 
\par\end{centering}

\caption{The differential conductance for an STM tunneling measurement on a
$p$-wave nanowire with $t=2\mu=10\Delta$ and $\gamma=0.001\Delta$.
In panels (a)-(d), the STM tip is placed at the end of the wire, where
a zero-energy Majorana mode is found. The results are computed under
the conditions of: (a) high resolution and low temperature, (b) high
resolution and high temperature, (c) low resolution and high temperature,
and (d) low resolution and high temperature. In panels (e)-(f), the
STM tip is placed at a site with an impurity, which induces a subgap
state with energy $0.3\Delta$, and the results shown correspond to
(e) high resolution and (f) low resolution, respectively. The high-temperature
results in (b,d,f) are very similar, indicating the difficulty in
distinguishing between states at zero energy and states at non-zero
but small energy in such circumstances; on the other hand, low-temperature
conductance curves with Majoranas in (a,c) can be clearly distinguished
from the curves in (e).\label{fig:Majorana}}
\end{figure}

We show in Fig.~\ref{fig:Majorana}(a)-(d) the conductance of a genuine
Majorana mode at the end of a clean $p$-wave nanowire, with no other
subgap states being present in the system. For comparison, in Fig.~\ref{fig:Majorana}(e)-(f)
we show the conductance of a subgap non-MF state induced by a non-magnetic
impurity in an otherwise uniform $p$-wave nanowire. We observe that
for low resolutions or high temperatures, the conductance spectra
for zero-energy Majorana modes cannot be effectively distinguished
from those for non-MF subgap modes. In particular, the high temperature
(and/or poor energy resolution) plots in Figs.~\ref{fig:Majorana}(e)
and (f) are indistinguishable from the corresponding plots in Figs.~\ref{fig:Majorana}(b)-(d),
thus confirming that high-temperature STM experiments, as carried
out in Ref.~\cite{Nadj-Perge2014Observation}, cannot really confirm
the existence of Majorana modes. Thus, only future experiments at
lower temperatures and employing better instrumental resolution would
be able to conclusively determine the existence or not of MFs in the
hybrid Fe nanowire-Pb superconductor system recently studied in Ref.~\cite{Nadj-Perge2014Observation}.

Our work (as well as that presented in Ref.~\cite{Dumitrescu2014Majorana})
shows that any subgap states, whether MF or not (and whether precisely
at zero-energy or not), would manifest very similar STM conductance
spectra at high experimental temperatures, and thus the observation
of broad, weak, and spectrally diffuse zero bias conductance {}``peak''
cannot be taken as being synonymous with the existence of MFs in the
system, as has been done rather uncritically in Ref.~\cite{Nadj-Perge2014Observation}.

\section{Conclusion}

In summary, we have established that the nanowire on superconductor
hybrid systems can potentially have very short Majorana localization
length even when the induced topological superconducting gap is very
small in the nanowire by virtue of the substrate induced strong renormalization
of the effective nanowire parameters (e.g. the Fermi velocity, the
gap, etc.) because of strong frequency dependence of the relevant
self-energy function determining the proximity-induced pair potential
in the nanowire. We have shown that this renormalization goes as $R^{-3}$
where $R$ is the effective nanowire confinement size in the transverse
direction determining how one-dimensional the system really is (with
R going to zero limit being the true 1D nanowire limit). This provides
an explanation for why the Majorana localization length could be very
small (large) in metallic (semiconducting) nanowires on superconductors
since $R\sim0.5(20)\,{\rm nm}$ in the two systems leads to a large
difference in the renormalization effect induced by the substrate.
The substrate-induced suppression of the Majorana localization length
may have implications for recent efforts \cite{Nadj-Perge2014Observation}
to observe localized Majorana modes in fairly short ($\lesssim 50$nm)
ferromagnetic 
Fe chains on superconducting Pb substrates using STM spectroscopy,
providing a possible explanation \cite{Peng2014Strong} for how the
Majorana mode may be spatially highly localized on a sub-nm length
scale near the ends of the Fe adatom chain in spite of a very small
induced superconducting gap.

But, the definitive observation of spatially localized Majorana fermions
(rather than just spatially localized ordinary fermionic subgap states)
requires precise energy localization (exactly at mid-gap or zero energy)
in addition to strong spatial localization. Such an {}``energy localization''
necessitates an experimental temperature much smaller than the topological
superconducting gap. This condition of {}``energy localization''
is absent in the experiment of Ref.~\cite{Nadj-Perge2014Observation}
since the temperature is larger than the estimated induced superconducting
gap. Thus, the spectral weight of the observed {}``zero bias'' conductance
feature is spectrally spread over the whole induced gap, making it
difficult to conclude about possible existence of Majorana fermions
in spite of the observed spatial localization. In addition, we have
shown through explicit numerical simulations that high-temperature
and low-resolution tunneling conductance measurements cannot distinguish
between Majorana modes and ordinary fermionic subgap states as both
manifest broad and weak zero-bias conductance features. The distinction
between these two situations necessitates experiments at temperatures
(and resolutions) well below the induced superconducting gap energy
in the nanowire. We mention in this context that the state of the
arts STM experiments on superconductors are routinely carried out
at temperatures of $100\,{\rm mK}$ or below
\cite{Firmo2013Evidence,*Allan2013Imaging,*ZhouNatPhys2013,HessPRL1990},
and future experiments with much lower temperatures (and much better
resolutions) than used in Ref.~\cite{Nadj-Perge2014Observation}
can settle the question of whether Majorana fermions have indeed been
observed or not in ferromagnetic chains on superconducting substrates.
The minimal existence proof of Majorana fermions necessitates the
demonstration of both spatial (at wire-ends) and spectral (at zero-energy)
localization of the observed excitation.

Lowering temperature (and/or enhancing the topological gap) as well
as improving instrumental resolution should lead to the MF-induced
ZBCP becoming sharper and stronger {[}as in Fig.~\ref{fig:Majorana}(a)-(d){]}
in longer ferromagnetic chains whereas in shorter chains, the ZBCP
should split because of the hybridization between the two Majorana
end modes \cite{Cheng2009Splitting,Cheng2010Stable,DasSarma2012Splitting}.
(The currently observed zero bias peak in the ferromagnetic chains
\cite{Nadj-Perge2014Observation} is only a very weak fraction ($10^{-4}$)
of the quantized value of $2e^{2}/h$ predicted for the Majorana zero
modes, and lowering of temperature should enhance its strength \cite{Sengupta2001Midgap}.)
On the other hand, if the observed sub-gap STM conductance features
are arising from impurity-induced non-MF subgap bound states {[}as
in Fig.~\ref{fig:subgap} or Fig.~\ref{fig:Majorana}(e)-(f){]},
then lowering temperature (and/or increasing the topological gap)
and enhancing resolution should clearly show that these accidental
subgap states are non-topological and therefore the resultant subgap
conductance features are not spectrally located exactly at zero energy.
There should not be any energy splitting of such non-MF subgap peaks
in shorter chains, also distinguishing them from possible MF peaks.
Current experiments \cite{Nadj-Perge2014Observation} do not actually
find any clear evidence for a topological gap in the ferromagnetic
chain with the background subgap conductance being typically more
than 50\% of the normal state conductance, and any definitive claim
for the observation of Majorana fermions must necessarily be coupled
with the observation of a reasonably well-defined topological gap
with background subgap conductance being suppressed by at least one
or two orders of magnitude from its current value. In this context,
it may be useful to remember that any fermionic subgap bound state
in the adatom (Fe)-superconductor (Pb) system will most likely also
be localized near the ends of the Fe chain since this is where the
impurity potential is the strongest, as was explicitly shown recently
\cite{Hui2015Majorana} for possible YSR states in this system.
Indeed a direct numerical diagonalization using model band structures
of the Fe/Pb experimental system produces a very large complex of
states in the superconducting gap \cite{Ji2008High-Resolution}, most
of which have nothing topological about them (but all of which are
likely to contribute to the subgap tunneling conductance), and the
complexity of the system therefore makes any straightforward
interpretation of the STM data
very difficult.
\begin{acknowledgments}
This work is supported by JQI-NSF-PFC and LPS-CMTC. 
\end{acknowledgments}
\appendix

\section{Derivation of Effective Hamiltonian}
\begin{widetext}
The various terms in the expression of the self-energy {[}Eq.~(\ref{eq:selfEnergyTerms}){]}
are

\begin{eqnarray}
Z & = & 1-\int\frac{d^{3}k}{\left(2\pi\right)^{3}}\frac{1}{\omega^{2}-\xi_{\bm{k}}^{2}-\Delta_{SC}^{2}}M_{1},\\
Zt_{\bm{r},\bm{r}'} & = & \int\frac{d^{3}k}{\left(2\pi\right)^{3}}e^{i\bm{k}\cdot\left(\bm{r}-\bm{r}'\right)}\frac{\xi_{\bm{k}}}{\omega^{2}-\xi_{\bm{k}}^{2}-\Delta_{SC}^{2}}M_{1},\\
Z\bm{\lambda}_{\bm{r},\bm{r}'}\cdot\hat{\bm{\sigma}} & = & \int\frac{d^{3}k}{\left(2\pi\right)^{3}}e^{i\bm{k}\cdot\left(\bm{r}-\bm{r}'\right)}\frac{\xi_{\bm{k}}}{\omega^{2}-\xi_{\bm{k}}^{2}-\Delta_{SC}^{2}}M_{2},\\
Z\Delta_{\bm{r},\bm{r}'} & = & -\int\frac{d^{3}k}{\left(2\pi\right)^{3}}e^{i\bm{k}\cdot\left(\bm{r}-\bm{r}'\right)}\frac{\Delta_{SC}}{\omega^{2}-\xi_{\bm{k}}^{2}-\Delta_{SC}^{2}}M_{1},\\
Z\bm{\Delta}_{\bm{r},\bm{r}'}\cdot\hat{\bm{\sigma}} & = & -\int\frac{d^{3}k}{\left(2\pi\right)^{3}}e^{i\bm{k}\cdot\left(\bm{r}-\bm{r}'\right)}\frac{\Delta_{SC}}{\omega^{2}-\xi_{\bm{k}}^{2}-\Delta_{SC}^{2}}M_{2},
\end{eqnarray}
 where $M_{1}\equiv t_{s}^{2}B_{s}^{T}B_{s}^{*}+t_{p}^{2}B_{p}^{T}B_{p}^{*}$
and $M_{2}\equiv t_{s}^{2}B_{s}^{T}B_{p}^{*}+t_{p}^{2}B_{p}^{T}B_{s}^{*}$.
If we take the specified form of $B_{s}$ and $B_{p}$ {[}Eq.~(\ref{Eq:BsBp}){]},
they become\begin{subequations} 
\begin{eqnarray}
M_{1}\left(\bm{k}\right) & = & t_{s}^{2}\cos^{2}\theta+t_{p}^{2}\sin^{2}\theta\\
M_{2}\left(\bm{k}\right) & = & t_{s}t_{p}\sin2\theta\hat{\bm{\sigma}}\cdot\bm{e}_{\bm{k}}
\end{eqnarray}
 \end{subequations}Further, restricting $\bm{r}$ only at the lattice
positions of the nanowire lying on the $x$-axis, we can evaluate
the terms exactly as\begin{subequations} 
\begin{eqnarray}
Z & = & 1-\int\frac{d^{3}k}{\left(2\pi\right)^{3}}\frac{1}{\omega^{2}-\xi_{\bm{k}}^{2}-\Delta_{SC}^{2}}M_{1}\left(\bm{k}\right)\nonumber \\
 & = & 1+\frac{\pi\nu}{\sqrt{\Delta_{SC}^{2}-\omega^{2}}}\left(t_{s}^{2}\cos^{2}\theta+t_{p}^{2}\sin^{2}\theta\right)\\
Zt_{n>0} & = & \int\frac{d^{3}k}{\left(2\pi\right)^{3}}e^{ik_{x}na_{{\rm lat}}}\frac{\xi_{\bm{k}}}{\omega^{2}-\xi_{\bm{k}}^{2}-\Delta_{SC}^{2}}M_{1}\left(\bm{k}\right)\nonumber \\
 & = & -\pi\nu\frac{\cos nk_{F}a_{{\rm lat}}}{nk_{F}a_{{\rm lat}}}\left(t_{s}^{2}\cos^{2}\theta+t_{p}^{2}\sin^{2}\theta\right)e^{-\frac{n\sqrt{\Delta_{SC}^{2}-\omega^{2}}a_{{\rm lat}}}{v_{F}}}\\
Z\bm{\lambda}_{n>0}\cdot\hat{\bm{\sigma}} & = & \int\frac{d^{3}k}{\left(2\pi\right)^{3}}e^{ik_{x}na_{{\rm lat}}}\frac{\xi_{\bm{k}}}{\omega^{2}-\xi_{\bm{k}}^{2}-\Delta_{SC}^{2}}M_{2}\left(\bm{k}\right)\nonumber \\
 & = & -i\pi\nu t_{s}t_{p}\sin2\theta\hat{\sigma}_{y}\frac{\cos nk_{F}a_{{\rm lat}}+nk_{F}a_{{\rm lat}}\sin nk_{F}a_{{\rm lat}}}{n^{2}k_{F}^{2}a_{{\rm lat}}^{2}}e^{-\frac{n\sqrt{\Delta_{SC}^{2}-\omega^{2}}a_{{\rm lat}}}{v_{F}}}\\
Z\Delta_{n} & = & -\int\frac{d^{3}k}{\left(2\pi\right)^{3}}e^{ik_{x}na_{{\rm lat}}}\frac{\Delta_{SC}}{\omega^{2}-\xi_{\bm{k}}^{2}-\Delta_{SC}^{2}}M_{1}\left(\bm{k}\right)\nonumber \\
 & = & \pi\nu\frac{\Delta_{SC}}{\sqrt{\Delta_{SC}^{2}-\omega^{2}}}\frac{\sin nk_{F}a_{{\rm lat}}}{nk_{F}a_{{\rm lat}}}\left(t_{s}^{2}\cos^{2}\theta+t_{p}^{2}\sin^{2}\theta\right)e^{-\frac{n\sqrt{\Delta_{SC}^{2}-\omega^{2}}a_{{\rm lat}}}{v_{F}}}\\
Z\bm{\Delta}_{n>0}^{(t)}\cdot\hat{\bm{\sigma}} & = & -\int\frac{d^{3}k}{\left(2\pi\right)^{3}}e^{ik_{x}na_{{\rm lat}}}\frac{\Delta_{SC}}{\omega^{2}-\xi_{\bm{k}}^{2}-\Delta_{SC}^{2}}M_{2}\left(\bm{k}\right)\nonumber \\
 & = & i\pi\nu t_{s}t_{p}\sin2\theta\hat{\sigma}_{y}\left(\frac{\Delta_{SC}}{\sqrt{\Delta_{SC}^{2}-\omega^{2}}}\frac{\sin nk_{F}a_{{\rm lat}}-nk_{F}a_{{\rm lat}}\cos nk_{F}a_{{\rm lat}}}{n^{2}k_{F}^{2}a_{{\rm lat}}^{2}}\right)e^{-\frac{n\sqrt{\Delta_{SC}^{2}-\omega^{2}}a_{{\rm lat}}}{v_{F}}}
\end{eqnarray}

The full effective BdG Hamiltonian of the nanowire is then given by
substituting these expressions to Eq.~(\ref{eq:Heff}).\end{subequations} 
\end{widetext}
\vfill{}

\section{Hamiltonian in the strong spin polarization limit}
\label{appendix:Heff}

To obtain the Hamiltonian in the limit of strong spin
  polarization, we start by writing Eq.~(\ref{eq:Heff}) in momentum space
\begin{eqnarray}
H_{{\rm eff}} & = &
\sum_{k}\left[(\epsilon_{k}-\mu)a_{k}^{\dagger}a_{k}+\lambda_{k}a_{k}^{\dagger}\sigma_{y}a_{k}+\tilde{B}
  a_{k}^{\dagger}\sigma_{z}a_{k}\right.\nonumber \\
 &  & \left.+\Delta_{k}^{\left(t\right)}\left(a_{k\uparrow}a_{-k\uparrow}+a_{k\downarrow}a_{-k\downarrow}+{\rm h.c.}\right)\right.\label{eq:H}\\
 &  & \left.+\Delta_{k}\left(a_{k\uparrow}a_{-k\downarrow}+{\rm h.c.}\right)\right],\nonumber 
\end{eqnarray}
where $a_{k}^{\dagger}=\left(a_{k\uparrow}^{\dagger},a_{k\downarrow}^{\dagger}\right)$,
$\epsilon_{k}$ is the Fourier transform of the hopping terms in
Eq.~(\ref{eq:Heff}), $\tilde{B} = Z^{-1}B$ is the renormalized
exchange field, while $\lambda_{k}$, $\Delta_{k}$, and
$\Delta^{(t)}_k$ are the proximity-induced spin-orbit coupling,
singlet pairing potential, and triplet pairing potential,
respectively. 
Diagonlizing
the first three terms in Eq.~(\ref{eq:H}) gives us the ``normal
state'' dispersion 
$\xi_{\pm}\left(k\right)=\epsilon_{k}-\mu\pm\sqrt{\tilde{B}^{2}+\tilde{\lambda}_{k}^{2}}$
with eigenvectors 
\beq
\phi_{\pm}\left(k\right)=\frac{1}{2}\left(1\pm\frac{\tilde{B}-i\lambda_{k}}{\sqrt{\tilde{B}^{2}+\lambda_{k}^{2}}},1\mp\frac{\tilde{B}-i\lambda_{k}}{\sqrt{\tilde{B}^{2}+\lambda_{k}^{2}}}\right)^{T}\,.
\eeq
Following the standard approach \cite{Lutchyn2010Majorana,Alicea2010Majorana},
we rewrite the Hamiltonian in this basis with the transformation
\begin{equation}
a_{k}=\phi_{+}\left(k\right)c_{+}\left(k\right)+\phi_{-}\left(k\right)c_{-}\left(k\right),
\end{equation}
where $c_{\pm}$ annihilates states in the corresponding normal state bands. Eq.~(\ref{eq:H})
then becomes\begin{widetext} 
\begin{eqnarray}
H_{{\rm eff}} & = & \sum_{k}\left\{ \xi_{+}\left(k\right)c_{+}^{\dagger}\left(k\right)c_{+}\left(k\right)+\xi_{-}\left(k\right)c_{-}^{\dagger}\left(k\right)c_{-}\left(k\right)+\Delta_{k}^{\left(t\right)}\left[c_{+}\left(k\right)c_{+}\left(-k\right)+c_{-}\left(k\right)c_{-}\left(-k\right)+{\rm h.c.}\right]\right.\nonumber \\
 &  &
\left.-\frac{i\lambda_{k}\Delta_{k}}{\sqrt{\tilde{B}^{2}+\lambda_{k}^{2}}}\left[c_{+}\left(k\right)c_{+}\left(-k\right)-c_{-}\left(k\right)c_{-}\left(-k\right)+ {\rm h.c.}\right]+\frac{\tilde{B}\Delta_{k}}{\sqrt{\tilde{B}^{2}+\lambda_{k}^{2}}}\left[c_{+}\left(k\right)c_{-}\left(-k\right)+{\rm h.c.}\right]\right\} .\label{eq:Hband}
\end{eqnarray}
\end{widetext}
The pairing terms on the first line correspond to the
proximity-induced triplet gap. On the second line, we have additional
triplet pairing terms induced from the interplay of the
proximity-induced spin-orbit coupling $\lambda_{k}$ and singlet gap
$\Delta_{k}$. In the strongly spin-polarized  limit where
$\tilde{B}\gg\left|\lambda_{k}\right|$, $|\Delta_k|$, however, these terms are
negligible. The singlet pairing term, also on the second line, is not
suppressed at the Hamiltonian level. However, in the strongly spin
polarized limit the band basis approximately coincides with
the spin basis, and so the singlet pairing potential does not open a
gap due to the  huge momenta mismatch between the 
two bands, and may thus be ignored in the effective Hamiltonian. 

\bibliography{MajLoc}

\end{document}